\documentclass[12pt]{article}
\usepackage[]{epsfig}
\usepackage{amsfonts}
\usepackage{amsmath}
\usepackage{color}
\usepackage{float}
\usepackage{multirow}

\newcommand{\be}{\begin{equation}}
\newcommand{\ee}{\end{equation}}
\newcommand{\ba}{\begin{eqnarray}}
\newcommand{\ea}{\end{eqnarray}}

 \newcommand{\bea}{\begin{eqnarray}} \newcommand{\eea}{\end{eqnarray}}

\newcommand{\cleqn}{\setcounter{equation}{0}}
\numberwithin{equation}{section}

\begin{document}

\title{\bf Holographic phase transitions from higgsed, non abelian charged black holes}

\author{
Gast\'on L. Giordano\footnote{gaston2031@gmail.com}\;,
Adri\'an R. Lugo\footnote{lugo@fisica.unlp.edu.ar}\\
{\normalsize \it Departamento de F\'\i sica - Universidad Nacional de La Plata}\\
{\normalsize \it and}\\
{\normalsize \it Instituto de F\'\i sica de La Plata IFLP - CONICET}\\
{\normalsize\it C.C. 67, 1900 La Plata, Argentina}
}
\date{\today}
\maketitle
\begin{abstract}
We find solutions of a gravity-Yang-Mills-Higgs theory in four dimensions that represent  asymptotic anti-de Sitter charged  black holes with partial/full  gauge symmetry breaking.
We then apply the AdS/CFT correspondence to study the strong coupling regime of a $2+1$ quantum field theory at temperature $T$ and finite chemical potential, which undergoes transitions to phases exhibiting the condensation
of a composite charged vector operator below a critical temperature $T_c$, presumably describing $p+ip/p$-wave superconductors.
In the case of $p+ip$-wave superconductors the transitions are always of second order.
But for $p$-wave superconductors we determine the existence of a critical value $\alpha_c$ of the gravitational coupling (for fixed Higgs v.e.v. parameter $\hat m_W$) beyond which the transitions become of first order.
As a by-product, we show that the $p$-wave phase is energetically favored over the $p+ip$ one, for any values of the parameters.
We also find the ground state solutions corresponding to zero temperature.
Such states are described by domain wall geometries that interpolate between $AdS_4$ spaces with different light velocities, and for a given $\hat m_{W}$, they exist below a critical value of the coupling.
The behavior of the order parameter as function of the gravitational coupling near the critical coupling suggests
the presence of second order quantum phase transitions.
We finally study the dependence of the solution on the Higgs coupling, and find the existence of a critical value beyond which no condensed solution is present.

\end{abstract}

\section{Introduction}
\cleqn

In recent years the application of AdS/CFT or more generally gauge/gravity correspondence \cite{Maldacena} \cite{GKP} \cite{Witten1998} to the study of condensed matter physics has attracted a lot of attention, providing in particular gravitational descriptions of systems exhibiting superconductor/superfluid phases \cite{HHH1} \cite{Hartnoll:2008kx}.
Since in condensed matter physics we are typically dealing with systems at finite charge density and temperature, in the context of the AdS/CFT correspondence the dual gravity descriptions should be given in terms of gravitational models with a negative cosmological constant which admit charged black holes as vacuum solutions.
In fact, a charged black hole naturally introduces a charge density/chemical potential and temperature in the quantum field theory (QFT) defined on the boundary using the gauge/gravity correspondence.
This set-up allows in particular to study phase transitions and construct phase diagrams in parameter space.

The simplest model is provided by an Einstein-Maxwell theory coupled to a charged scalar field that,
in the framework of the AdS/CFT correspondence, is dual to a scalar operator which carries the charge
of a global $U(1)$ symmetry.
It has been shown that a charged black hole solution, interpreted as the uncondensed phase, becomes unstable and develops scalar hair at low temperature breaking the $U(1)$ symmetry near the black hole horizon \cite{Gubser:2008px} \cite{Gubser1}.
This phenomenon in general may be interpreted as a second order phase transition between conductor and superconductor phases, interpretation that is supported by analyzing the behavior of the conductivity in these phases \cite{Hartnoll:2008kx}.
There were also studied vortex like solutions that describe type II holographic superconductors \cite{AJ} \cite{MPS} \cite{Domenech:2010nf} and more recently, spatially anisotropic, abelian models of superconductors \cite{Koga:2014hwa}.

Soon after these ``$s$-wave" holographic superconductor models were introduced, holographic superconductors models with vector hair, known as $p$-wave holographic superconductors, were explored numerically first in \cite{GubserNonAb} and \cite{GubserP} (for a recent analytical treatment, see \cite{Gangopadhyay:2012gx}).
The simplest example of $p$-wave holographic superconductors may be provided by an Einstein-Yang-Mills theory with $SU(2)$ gauge group and no scalar fields, where the electromagnetic gauge symmetry is identified with an $U(1)$ subgroup of $SU(2)$.
The other components of the $SU(2)$ gauge field play the role of charged fields
dual to some vector operators whose non-zero expectation values break the $U(1)$ symmetry leading to a phase transition in the dual field theory.

More recently, solutions to gravity-matter field equations where both scalar and vector order parameters are present were considered; they describe systems where competition/coexistence of different phases takes place \cite{Nie:2013sda}-\cite{Nie:2014qma}.

Regular, self-gravitating dyonic solutions of the Einstein-Yang-Mills-Higgs (EYMH) equations in the BPS limit
and asymptotic to global AdS space, were constructed time ago in \cite{LS} and \cite{LMS}.
They were extended to dyonic black hole solutions in \cite{LMS1} and \cite{LMS2},
where they were interpreted as describing a so-called $p + ip$-wave superconductor (isotropic) system at finite temperature in the condensed phase.
The purpose of the present work is to generalize previous results by finding more general black hole solutions of EYMH in asymptotically $AdS_4$ space  with finite mass and electric charge density, to interpret them via the gauge/gravity duality as describing phases of a strongly coupled field theory, and to construct the corresponding phase diagrams
\footnote{
We will be considering the usual plane horizon ansatz,
relevant to study condensed matter systems with translational invariance.
Under these circumstances, the magnetic charge density of the dyon solutions in  \cite{LMS1} \cite{LMS2} disappears.
}.
More specifically, in first term we start by revisiting the analysis of \cite{LMS1} \cite{LMS2},
verifying the existence of second order phase transitions all along the parameter space.
It was found (see for example \cite{Ammon:2009xh}) that some holographic systems pass from a second order phase transition as a function of the temperature in the non back-reaction limit to a first order when the gravitational coupling exceeds a certain value.
Such a phenomenon occurs in holographic superfluids when the velocity is high enough \cite{Basu:2008st} \cite{Herzog:2008he}, and it was measured in certain types of superconductors \cite{superconductor0orden} \cite{superconductor01orden} \cite{superconductor1orden}.
We have found this kind of behavior in our system in the anisotropic case, finding condensed solutions and constructing the phase diagram.
Second, we compute free energies and find that for any set of values of the free parameters that determines the solutions, the anisotropic phase is energetically favored over the isotropic phase, as conjectured in other contexts \cite{Basu:2009vv} \cite{Roberts:2008ns} \cite{Nacho}.
Third, we analyze the zero temperature limit, case that had not been addressed before; for low enough gravitational coupling we find solutions which spontaneously break the $U(1)$ symmetry and have zero entropy, and so describe the true ground state of the system.
For gravitational couplings higher than a critical value the solution disappears, which is interpreted as a
second order quantum phase transition.
Lastly, we study the effect of a non zero Higgs potential on the system.

The paper is organized as follows.
In section $2$ we present the model and write the translational invariant ansatz for the fields and the equations of motion that reduces to a a nonlinear system
of coupled ordinary differential equations.
In section $3$ we present generalities of the systems to be studied at non-zero temperature, in particular the analysis of the holographic map to be used.
In section $4$ we present the numerical results concerning the ``BPS limit", i.e. null Higgs potential, including computations of free energies.
Section $5$ is devoted to the study of the zero temperature case and the description of the ground state of the superconductor, including the presence of quantum phase transitions as function of the gravitational coupling and variable Higgs vacuum expectation value (vev) $\hat m_{W}$.
In section $6$ the effect of a non-zero Higgs potential is considered.
A summary and discussion of the results is given in section $7$.
Finally two appendices are added, one containing the boundary expansions of the fields and other containing the
equations of motion and free energy in other parameterization commonly used in the literature.

\section{The gravity-Yang-Mills-Higgs system}
\cleqn

\subsection{The model}

We consider a gravity-Yang-Mills-Higgs system in a $1+3$ dimensional space-time
with Minkowski signature $(-+++)$.
We take $SU(2)$ as the gauge group, with generators satisfying the algebra,
\be
[X_a , X_b] = \epsilon_{abc}\; X_c\qquad;\qquad a,b,c=0,1,2\quad,\quad \epsilon_{012}\equiv +1
\ee
and the scalar field in the adjoint representation, $H = H^a\,X_a$.
The full action to be considered is,
\be
S = S^{(bulk)} + S^{(GH)} +S^{(ct)}
\ee
where
\bea\label{S}
S^{(bulk)} &=& \int_{\cal M} d^4x\,\sqrt{|g|}\; \left(\frac{1}{2\,\kappa^2}  \left(  R + \frac{6}{L^2} \right)-\frac{1}{4\,e{}^2}\; F_{MN}^a F^{a\,MN}\right.\cr
&-&\left.\frac{1}{2}\,D^M H^a \; D_M H^a-
\frac{\lambda}{4}\; ( H^a H^a - H_0{}^2 )^2 \right)\cr
S^{(GH)}&=&\frac{1}{2\,\kappa^2}\;\int_{\partial{\cal M}} d^3x\,\sqrt{|h|}\;2\,K
\eea
where  $\kappa$, $e$ and $\lambda$ are the gravitational, gauge and scalar couplings respectively, $L$ is the AdS scale related to the negative cosmological constant through $\Lambda=-3/L^2$, and $H_0>0$ defines the vacuum expectation value of the Higgs field (and so the boundary condition at infinity, see below in (\ref{bcinfty})).
As it is well-known, the Gibbons-Hawking term $S^{(GH)}$ is necessary to have a well defined variational principle \cite{Wald}, where $K\equiv\nabla_a n^a$ is the trace of the extrinsic curvature, and $h$ and $n$ the induced metric and normal vector on $\partial{\cal M}$.
The counter-term action $S^{(ct)}$ will be discussed in section \ref{numersn}.

The field strength $F^a_{MN}$  and  the covariant derivative $D_M$ acting on the
Higgs triplet $H^a$ are defined as,
\be\label{strenght}
F^a_{MN}\equiv\partial_M A_N^a-\partial_N A_M^a+\epsilon_{abc}\;A_M^b\; A_N^c \quad;\quad
D_M H^a\equiv\partial_M H^a + \epsilon_{abc}\; A_M^b\; H^c
\ee

Let us consider coordinates $(x^\mu, y)$ and an ansatz preserving translational invariance in the coordinates $\{x^\mu, \mu=0,1,2\}$,
\bea\label{gralansatz}
g &=& - f(y)\; A(y)^2\; {dx^0}^2 + y^2\left(c(y)^2\;{dx^1}^2  +{dx^2}^2\right)
+ L^2\;\frac{ dy^2}{f (y)}\cr
A &=& L^{-1}\left(dx^0\;J(y)\;X_0+ dx^1\;K_1(y)\;X_1  + dx^2\;K_2(y)\;X_2\right)\cr
H &=& H_0\; H(y)\; X_0
\eea
In what follows it will be convenient to introduce the dimensionless coupling constants,
\be\label{dlesspar}
\alpha\equiv \frac{\kappa}{e\,L}\qquad;\qquad \hat m_W \equiv e\, H_0\,L
\qquad;\qquad\lambda_0 \equiv e^2\,H_0{}^4\,L^4\,\lambda
\ee

The gravity equations of motion (e.o.m.) derived from (\ref{S}) result,
\bea\label{graeom1}
-\left( y\; f (y)\right)' + 3\,y^2 &-& y^2\,f(y)\,\frac{c''(y)}{c(y)} -
\left(3\,y\,f(y)+\frac{y^2}{2}\,f'(y)\right)\;\frac{c'(y)}{c(y)}\cr
&=&\alpha^2\;\left( \frac{\lambda_0}{4}\;y^2\;(H(y)^2 -1)^2 +
f(y)\;V_1 + V_2 + \frac{y^2}{2}\;\frac{J'(y)^2}{A(y)^2}\right.\cr
&+&\left.\frac{1}{2}\left(\frac{K_1(y)^2}{c(y)^2}+ K_2(y)^2\right)\;
\left( \hat m_W{}^2\,H(y)^2 + \frac{J(y)^2}{f(y)\,A(y)^2}\right)\right)\cr
y\;\frac{A'(y)}{A(y)} - \frac{A(y)}{2\,c(y)}\;\left(\frac{y^2\,c'(y)}{A(y)}\right)'
&=&\alpha^2\;\left(V_1+\frac{1}{2}\,\left(\frac{K_1(y)^2}{c(y)^2}+ K_2(y)^2\right)\;\frac{J(y)^2}{ f(y)^2\, A(y)^2}\right)\cr
\frac{1}{A(y)\,c(y)}\;\left(y^2\,f(y)\,A(y)\,c'(y)\right)'&=&
\alpha^2\;\left(\left(\frac{K_1(y)^2}{c(y)^2}- K_2(y)^2\right)\;
\left(\frac{J(y)^2}{f(y)\,A(y)^2} - \hat m_W{}^2\,H(y)^2\right)\right.\cr
&-&\left.f(y)\;\left(\frac{K'_1(y)^2}{c(y)^2}- K'_2(y)^2\right)\right)
\eea
while that the matter e.o.m. are,
\bea\label{matteom1}
\frac{c(y)}{A(y)}\,\left(\frac{f(y)\,A(y)}{c(y)}\, K_1'(y) \right)'
&=& \left( \frac{ K_2(y)^2}{y^2} +\hat m_W{}^2\; H(y)^2  -
\frac{J(y)^2}{f(y)\, A(y)^2} \right)\;K_1(y)\cr
\frac{1}{A(y)\,c(y)}\,\left(f(y)\,A(y)\,c(y)\, K_2'(y) \right)'
&=& \left( \frac{ K_1(y)^2}{c(y)^2\,y^2} +\hat m_W{}^2\; H(y)^2  -
\frac{J(y)^2}{f(y)\, A(y)^2} \right)\;K_2(y)\cr
\frac{1}{A(y)\,c(y)}\left( y^2\,f(y)\,A(y)\,c(y)\,H'(y) \right)'&=&
\left(\frac{K_1(y)^2}{c(y)^2}+ K_2(y)^2+ \frac{\lambda_0}{\hat m_W{}^2}\,y^2\,(H(y)^2-1)\right)\;H(y)\cr
\frac{f(y)\;A(y)}{c(y)}\;\left( \frac{y^2\,c(y)}{A(y)}\,J'(y)\right)' &=& \left(\frac{K_1(y)^2}{c(y)^2}+ K_2(y)^2\right)\;J(y)
\eea
where we have defined,
\be\label{poten}
V_1 = \frac{1}{2}\;\left(\frac{K'_1(y)^2}{c(y)^2}+K'_2(y)^2\right)+
\frac{\hat m_W{}^2}{2}\;y^2\;H'(y)^2\qquad;\qquad
V_2 = \frac{1}{2}\;\frac{K_1(y)^2\,K_2(y)^2}{y^2\,c(y)^2}
\ee

We will start by considering the ``BPS limit" $\;\lambda_0 = 0$,
but conserving the crucial Higgs vacuum value $H_0>0$.
The effect of a finite Higgs coupling will be considered in section $6$.

\subsection{Boundary conditions}

We will search for charged black hole solutions which present
a horizon at $y=y_h$ where $f(y_h)=0$.
The associated Bekenstein-Hawking temperature of the black hole is given by,
\be\label{Tbh}
T_{BH}=\frac{1}{4\pi L}\,A(y_{h})\,f^{\prime}(y_{h})
\ee
The ansatz (and e.o.m.) are invariant under the scale transformations,
\bea\label{si1}
(x^0; A(y), J(y))&\longrightarrow& \left(\frac{x^0}{\beta}; \beta\;A(y),\beta\;J(y))\right)\cr
(x^1; c(y), K_1(y))&\longrightarrow& \left(\frac{x^1}{\beta'};\beta'\;c(y),\beta'\; K_1(y)\right)
\eea
They allow to fix some normalization imposing the b.c.,
$\;A(y), c(y)\stackrel{y\rightarrow\infty}{\longrightarrow} 1$,
in such a way that the $x^\mu$'s are identified with the minkowskian coordinates of the boundary QFT, and (\ref{Tbh}) with its temperature.
Furthermore there exists another scaling symmetry,
\be\label{si2}
\left(x^\mu, y\right)\rightarrow\left(\frac{x^\mu}{\gamma},\gamma\,y\right)\quad,\quad
f(y)\rightarrow\gamma^2\,f(y)\quad,\quad
K_i(y)\rightarrow\gamma\, K_i(y)\quad,\quad
J(y)\rightarrow\,\gamma\,J(y)
\ee
that if $y_h\neq 0$, allows to fix $y_h =1$
\footnote{
This is the case except when we consider the zero temperature limit.
When back-reaction is not taking into account $y_h=0$ corresponds to AdS space;
when it is considered, $y_h=0$ is imposed in order to get a true ground state description and (\ref{si2})
can be used to fix the chemical potential, see section \ref{T=0}.
}.
Since now on we will fix the position of the horizon in this way,
having in mind that we have to consider only scale invariants quantities.


In \cite{LMS1} and \cite{LMS2} solutions to (\ref{graeom1})-(\ref{matteom1})
with a horizon and asymptotically $AdS_4$ were studied.
More specifically, there were found solutions with $K_1=K_2=K$ and the following boundary conditions;
near the horizon $y\rightarrow 1^+$,
\bea\label{bchor}
f(y)&=& f_1\,(y-1) + {\cal O}[(y - 1)^2]\cr
A(y)&=& a_0 + a_1\,(y-1) + {\cal O}[(y - 1)^2]\cr
c(y)&=& c_0 + c_1\,(y - 1) + {\cal O}[(y - 1)^2]\cr
H(y)&=&  h_0 + h_1\,(y - 1) +{\cal O}[(y - 1)^2]\cr
K(y)&=&  k_0 + k_1\,(y - 1) +{\cal O}[(y - 1)^2]\cr
J(y)&=&  j_1\,(y - 1) + {\cal O}[(y - 1)^2]
\eea
while on the boundary $y\rightarrow\infty$,
\bea\label{bcinfty}
f(y)&=& y^2 +\frac{F_{1}}y + \cdots\cr
A(y)&=& 1 + \cdots\cr
c(y)&=& 1 +\cdots\cr
H(y)&=& 1 +\frac{H_1}{y^3} + \cdots\cr
K(y)&=& \frac{K_1}{y^{\kappa_1}} + \cdots\cr
J(y)&=& J_0 +\frac{J_1}{y} + \cdots
\eea
where consistency with the e.o.m. and finiteness of $K(y)$ fixes $\kappa_1$ to be,
\be\label{kappa1}
\kappa_1\;(\kappa_1 -1)=\hat m_W{}^2\qquad\longrightarrow\qquad
\kappa_1 = \frac{1}{2} + \sqrt{\frac{1}{4} + \hat m_W{}^2}
\ee
For more about the b.c. at the boundary, we refer the reader to the appendix A.
We will adopt the b.c. (\ref{bchor})-(\ref{bcinfty}) in this paper except in section \ref{T=0} where the b.c. on the horizon will have to be modified.

The bulk theory is invariant under the gauge group $SU(2)$; however the b.c. on the Higgs field, $\;H(y)\stackrel{y\rightarrow\infty}{\longrightarrow} 1\;$, breaks this invariance to the $U(1)$ generated by $X_0$.
With respect to this gauge subgroup the electric charge density of a solution is defined as usual by,
\be\label{denscarga}
\rho\equiv\frac{1}{V_2}\;\int_{\Re^2}\;*F|_{U(1)} = \frac{1}{L^2}\;\frac{c(y)}{A(y)}\;y^2\;J'(y)|_{y\rightarrow\infty}= -\frac{J_1}{L^2}
\ee
As we show in section \ref{numersn}, at fixed couplings $(\alpha, \hat m_W)\,$
a general solution to (\ref{graeom1})-(\ref{matteom1}) with the b.c. (\ref{bchor})-(\ref{bcinfty}) is determined by $J_0$, which is related to the $U(1)$ chemical potential by,
\be\label{potquim}
\mu \equiv A_0^0(\infty) = \frac{J_0}{L}
\ee
From (\ref{denscarga}) and (\ref{potquim}) the standard asymptotic expansion
follows,
\be
A_0^0(y) = \mu - \frac{L\,\rho}{y}+ \dots
\ee
Along this paper we will adopt $\mu$ as our scale.
From (\ref{si2}) the dimensionless, scale invariant  temperature is,
\be
T\equiv\frac{T_{BH}}{\mu} = \frac{a_0\,f_1}{4\,\pi\,J_0}
\ee
where $a_0$ and $f_1$ are defined in (\ref{bchor}).
A solution is determined by the three free parameters ($\alpha, \hat m_W , J_0$), and so the temperature
(through the coefficients $a_0$, $f_1$) results a function of them
\footnote{
In EYM systems where the Higgs field is not present the temperature is function of the only free parameter of the theory, $\alpha$, see \cite{Ammon:2009xh}.
}.

In the analytic solution to the equations (\ref{graeom1})-(\ref{matteom1}) that  preserves the $U(1)_{X_0}$ symmetry matter fields take the form,
\be\label{uncond1}
J(y) = J_0 + \frac{J_1}{y} =  J_0\,\left(1- \frac{1}{y}\right)\qquad;\qquad K_i(y) = 0\qquad;\qquad H(y) = 1
\ee
where we imposed smooth behavior of the gauge field at the horizon which yields the condition $J(1) = 0$, see the last line in (\ref{bchor}), and then fixed
$J_1 = - J_0$
\footnote{
When $y_h=0$, $J(y)=J_0$ is just the chemical potential and the metric  solution is AdS space; it describes the uncondensed phase when the temperature is zero, see section $5$.
}.
In what the metric functions concern, they correspond to the AdS Reissner-Nordstr\"om (AdS-RN) black hole,
\bea\label{uncond2}
A(y)&=& c(y) = 1\cr
f(y)&=& y^2 - \left(1+\frac{\alpha^2\,J_0{}^2}{2}\right)\,\frac{1}{y}+\frac{\alpha^2 J_0{}^2}{2}\,\frac{1}{y^2}\cr
&=& \frac{y-1}{y^2}\,\left(3 - \frac{\alpha^2 J_0{}^2}{2} + (y^2 + 2\,y + 3)\,(y-1)\right)
\eea
with temperature,
\be\label{uncond3}
T = \frac{1}{4\,\pi\,J_0}\,\left(3 - \frac{\alpha^2 J_0{}^2}{2}\right)
\ee
The extremal, zero temperature AdS-RN black hole is defined by the relation $\alpha^2 J_0{}^2 =6$.

\section{Solutions at $T>0$: superconducting state}

When the ``magnetic part" of the gauge field is non-trivial, i.e. $K_i(y)\neq 0$ for some $i=1,2$, the solution breaks not only the $U(1)_{X_0}$ invariance, but also the invariance under rotations in the $(x^1, x^2)$-plane.
According to the AdS/CFT dictionary this hair is interpreted as a spontaneous breaking
of a global $U(1)$ symmetry present in the boundary QFT, whose currents take an  expectation value,
\be\label{Jvev}
\langle J^a_i(x)\rangle\sim K_i\,\delta_i^a\qquad;\qquad i,a=1,2
\ee
Giving that the order parameter is dual to (a component of) the gauge field we are presumably modeling a $p$-wave superconductor \cite{GubserP}.
The normal state of the superconductor is described by the AdS-RN solution
(\ref{uncond1})-(\ref{uncond2});  such solution is energetically favored
until a critical temperature $T_{c}$ is reached; when $T<T_{c}$ the non  symmetric, hairy solution gives rise to a superconductor phase.

We remark that with the b.c. on the Higgs field we are breaking explicitly
the gauge group from $SU(2)$ to $U(1)_{X_0}$;
this yields a mass for the ``W" gauge bosons,
\be
m_W\equiv e\,H_0
\ee
The problem is thus the following: can we find under this condition a solution with $K_i(y)\neq 0$ that breaks spontaneously the $U(1)_{X_0}$?
In the boundary QFT  this is then interpreted as the breaking of a global $U(1)$ symmetry as it happens in superfluids and superconductors with weakly coupled photons.
From here we identify $T_c$  with the critical temperature of the phase transition in the QFT.

We will consider two cases.

\begin{itemize}

\item\underline{The isotropic case: $K(y)\equiv K_1(y)=K_2(y)$}

Although both gauge and rotational symmetries are broken by a hairy solution,
a configuration $(\ref{gralansatz})$ with $K_1(y)=K_2(y)$ preserves the diagonal subgroup,
$\\\left(U(1)_{X_0}\times SO(2)_{rot}\right)_{diag}$, fact that is manifest in (\ref{Jvev}) \cite{GubserNonAb}.
This configuration give rise to an energy-momentum tensor isotropic in the $x^1$-$x^2$ plane; therefore the metric function $c(y)$ must be a constant, even when back-reaction is taken into account.

This kind of configurations were first studied in \cite{LMS1}-\cite{LMS2}, using relaxation methods.
We will re-obtain these solutions here for later use by using shooting methods.

\item\underline{ The anisotropic case: $K(y)\equiv K_1(y)\;; \,K_2(y)=0$}

As stated above, a configuration with $K_1=0$ preserves the $U(1)_{X_0}$ and spatial rotations.
When $K_{1}(y)$ develops a non zero value the gauge symmetry $U(1)_{X_0}$ breaks, and the condensate  $K_{1}(y)\,X_1\,dx^1$ choose a direction $x^{1}$ as a special one.
Then if we take into account back-reaction effects the system cannot support the condition
$\;g_{11} = g_{22}$ \cite{GubserP}.
Due to this fact $\;T_{x^{1}x^{1}}\neq T_{x^{2}x^{2}}$ and the function $c(y)$ can not be a constant; in conclusion the system will be in an anisotropic phase.
\end{itemize}

In both cases the vacuum expectation value in the $d=3$ field theory of the current operator ${\cal O}_K$, dual to the function $K$ associated with the magnetic field in the bulk, follows from the identification,
$\langle O_K \rangle \sim K_1 $  with  $K_1$ defined in (\ref{bcinfty});
$K_1 = K_1(T)$ can be taken as the order parameter
that describes the phase transition of the system.
As discussed for different models \cite{Gubser1}-\cite{LMS2} one can interpret
this result by stating that a condensate is formed  above a black hole horizon
because of a balance of gravitational and electrostatic forces.
From the asymptotic behavior in  (\ref{bcinfty}) we get the dimension $\Delta[O_K]$  of the operator $\, O_K$ \cite{Aharony}
\be\label{dimOP}
\Delta [O_K] = 1 + \kappa_1 = \frac32 +\frac12\sqrt{1 + 4\,\hat m_W{}^2}
\ee

From numerical solutions we conclude that a finite temperature continuous symmetry breaking transition takes place so that the system condenses  at a critical temperature $T_c$, as can be seen from the behavior of
$K_1(T)$ for $T \approx T_c$ in figures $3$, $4$ and $5$.
Furthermore, we compare the free energies corresponding to both phases in figures $8$ and $9$, finding that the anisotropic phase is favored, see \cite{Basu:2009vv} \cite{Roberts:2008ns} \cite{Nacho} for related results.

\section{Numerical Solutions}\label{numersn}
\cleqn

We analyzed numerically equations (\ref{graeom1})-(\ref{matteom1}) and found solutions that satisfy the required b.c. (\ref{bchor})-(\ref{bcinfty}) in a wide region of the parameter space, that lead to the phase diagram in figure $1$.
Such solutions in the anisotropic case are shown in figure $2$.

Before presenting the results, we think is worth to spend a few words on the method used.
As discussed in appendix A, after fixing some normalization and asking for finiteness the solution near the boundary admits the expansions in equation (\ref{bdryexp}), and is determined by six constants,
$\,(F_1, C_1, J_0, J_1, K_1, H_1)$.
However the b.c.  on the horizon impose five conditions.
The first two come from the definition of the horizon and the regularity of the gauge field,
\be\label{conshor1}
f(1)=0\qquad;\qquad J(1)=0
\ee
They essentially fix the mass ($\sim F_1 $) and the charge density ($\sim J_1$) of the black hole.
The remaining three conditions fix $(C_1, K_1, H_1)$ and are obtained from an analysis of the (singular) behavior of the e.o.m. near the horizon,
\bea\label{conshor2}
c'(1)&=&\alpha^2\,\hat m_W{}^2\,\frac{c(1)}{f'(1)}\left(-\frac{K_1(1){}^2}{c(1)^2}+K_2(1)^2
\right)\, H(1){}^2\cr
K'_{1}(1)&=&\frac{K_{1}(1)}{f^{\prime}(1)}\left(K_2{}(1)^2 + \hat m_W{}^2\;H(1)^2\right)\cr
H'(1)&=&\frac{H(1)}{f'(1)}\,\left(\frac{K_1{}(1)^2}{c(1)^2}+K_2(1)^2\right)
\eea
Therefore the only additional free parameter that determines the solution is $J_0$, i.e. the chemical potential (\ref{potquim}).
In practice we integrate the system from the horizon, where according to (\ref{conshor1})-(\ref{conshor2}) the free parameters are,
\begin{equation}
J^{\prime}(1)=j_{1}\quad;\quad K(1)=k_{0} \quad;\quad H(1)=h_{0}\quad;\quad A(1)=a_{0}\quad;\quad c(1)=c_{0}
\end{equation}
as defined in (\ref{bchor}).
These parameters are selected in such a way that the solution matches the conditions on the boundary $(\ref{bcinfty})$,
\begin{equation}\label{bcdefinitivas}
A(\infty)=c(\infty)=H(\infty)=1\qquad;\qquad K(\infty)=0\qquad;\qquad
J(\infty)=J_0
\end{equation}.

\begin{figure}[H]\label{dfases}
\begin{center}
\includegraphics[height=5cm,width=5cm]{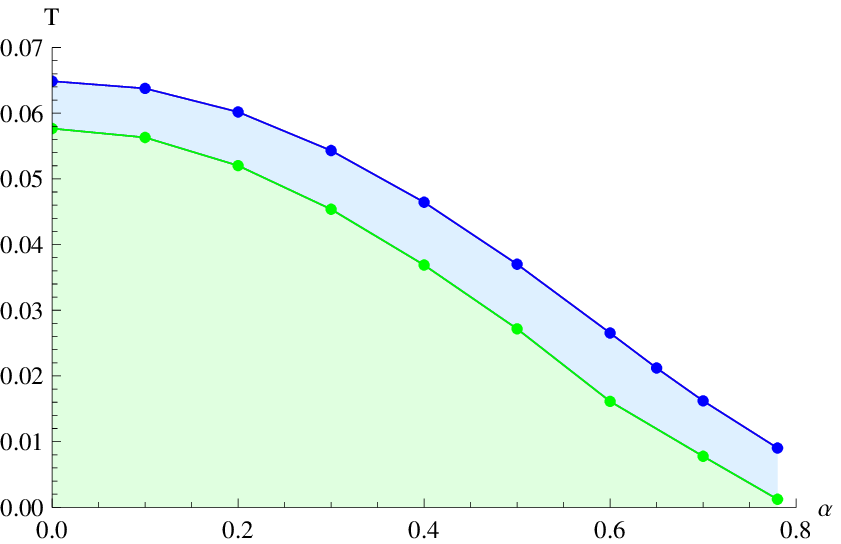}
\includegraphics[height=5cm,width=5cm]{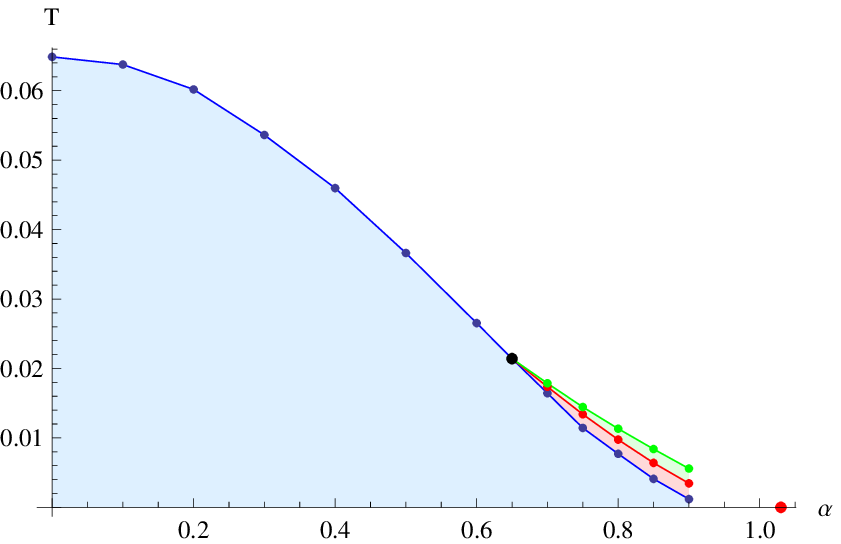}
\includegraphics[height=5cm,width=5cm]{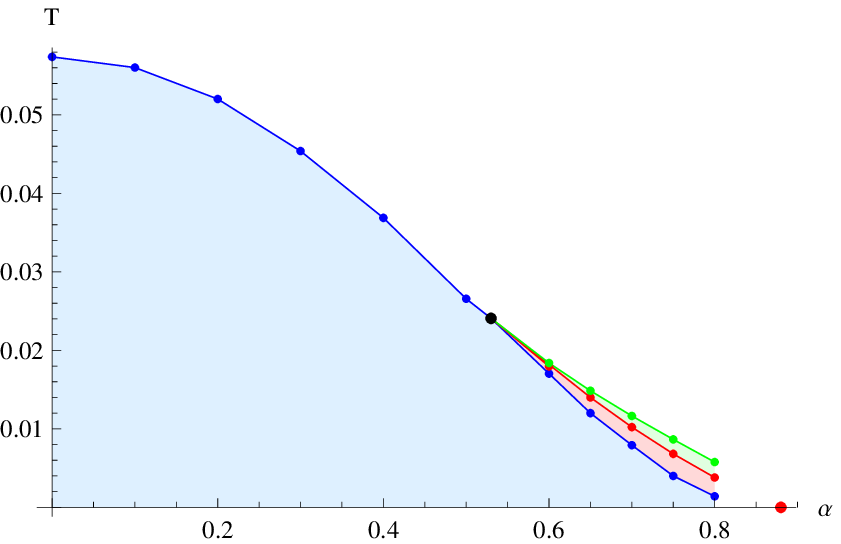}
\caption{Phase diagrams in the isotropic case (left) for $\hat m_{W} = 0.1$ (blue), $\hat m_{W} = 0.4$ (green), and
in the anisotropic case, for $\hat m_{W} = 0.1$ (center) and $\hat m_{W} = 0.4$ (right).
}
\end{center}
\end{figure}

Figure $1$ displays the phase diagrams in the $\alpha-T$ plane, for two different values of $\hat m_W$.
In the white regions only the normal or uncondensed phase is present.
In the isotropic case the system experiments second order phase transitions along the blue ($\hat m_{W} = 0.1$)
and green ($\hat m_{W} = 0.4$) curves.
In the anisotropic case, in the blue and red regions the condensed phase is the thermodynamically preferred phase.
The blue line until the black point indicates a critical line of second order transitions.
The black point signals the coupling $\alpha_{tc}$ beyond which the transitions become of first order along
the red line, while that the blue and green lines that continue after this tri-critical point represent spinodal lines.
The critical red curve of first order phase transitions ends in the red point at $T=0$, which represents a quantum phase
transition, signaling a critical coupling $\tilde\alpha$ above which the condensed phase ceases to exist, see Section  \ref{T=0}.
By comparing both graphs we can see that both $\alpha_{tc}$ and $\tilde\alpha$ decrease with increasing $\hat m_{W}$.
Similar phase diagrams were obtained in reference \cite{Erdmenger:2012ik} in absence of Higgs fields.

In figure $2$ the fields are shown as functions of the coordinate $y$, at fixed $J_0$ and $\hat m_W$ and for different $\alpha$'s.
For $\alpha_c\approx 0.8825$ a second horizon appears, as displayed from the curves corresponding to $f(y)/y^2$.
The uncondensed and condensed phases are separated by a curve on which the formation of the second horizon takes place for a given critical temperature determined by the gauge boson mass $\hat m_W$ and $J_0$.
In the isotropic case the curve is displayed in figure $1$ for two different values of $\hat m_W$
(blues and green lines) and it coincides with the critical curve on which the phase transitions take place.
On the other hand, in the anisotropic case the curve coincides with the critical curve (blue line in figure $1$)
until the tri-critical point $\alpha_{tc}$, and it continues through the spinodal curve in green.

In figure $3$ the order parameter $K_1$ in the isotropic case is plotted as a function of the temperature for different values of $\hat m_W$, at fixed gravitational coupling $\alpha=0.7$.
In this case the transition is of second order independently of $\alpha$, in agreement with \cite{GubserNonAb}.

Figures $4$ and $5$ shows the order parameter $K_1$ as function of $T$ in the anisotropic case from two perspectives:
at fixed $\hat m_W=0.4$ and varying $\alpha$ in figure $4$ and at fixed $\alpha=0.7$ and varying $\hat m_W$
in figure $5$.
From figure $4$ it is seen that for $\alpha_{tc}\approx 0.53$ $K_1$ becomes multi-valued, fact that signals the passage from second to first order phase transitions as corroborated from the free energy computations of the next subsection.
This phenomenon has been found recently in $p$-wave superfluids by studying the role of the back-reaction in the phase transitions \cite{Ammon:2009xh} (for experimental results on first order phase transitions in superconductors,
see \cite{superconductor0orden}\cite{superconductor01orden}\cite{superconductor1orden}).
By comparing figures $3$ and $5$ it is observed that the temperature at which the order parameter becomes zero is the same in both cases, and that the critical temperature decreases when $\hat m_W$ increases,
what can be interpreted as the presence of the Higgs field hinders the condensation.
Furthermore, we have checked that near $T_{c}$ and for weak gravitational couplings $\alpha<\alpha_{tc}$, $K_1$ behaves like $(T_c-T)^\frac{1}{2}$, indicating a second order phase transition with mean field exponent $\frac{1}{2}$
as usually happens in holographic descriptions of critical systems in the limit of large number degrees of freedom.

\begin{figure}[H]\label{campos}
\begin{center}
\includegraphics[height=5cm,width=7.5cm]{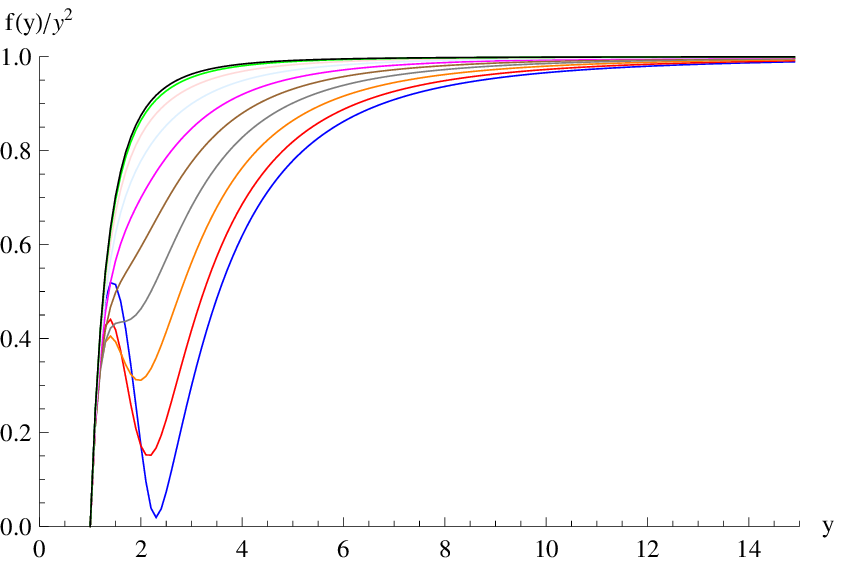}
\includegraphics[height=5cm,width=7.5cm]{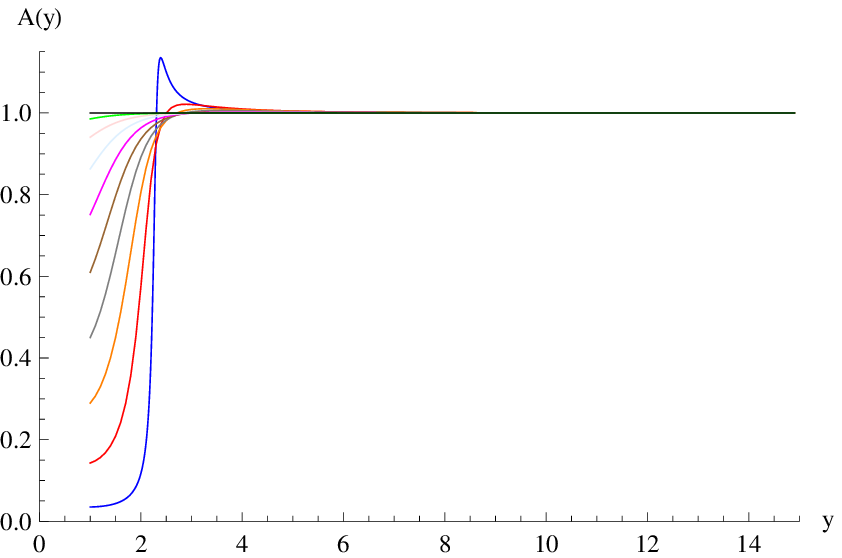}\\
\includegraphics[height=5cm,width=7.5cm]{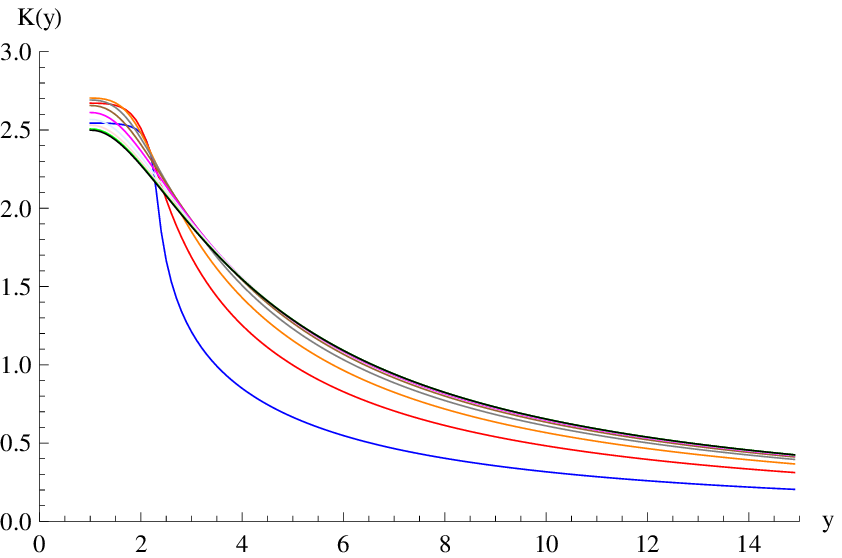}
\includegraphics[height=5cm,width=7.5cm]{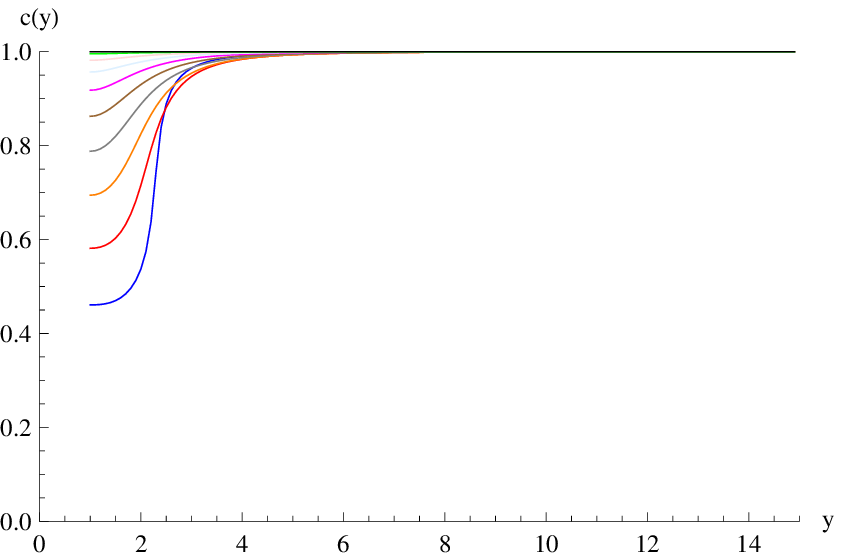}\\
\includegraphics[height=5cm,width=7.5cm]{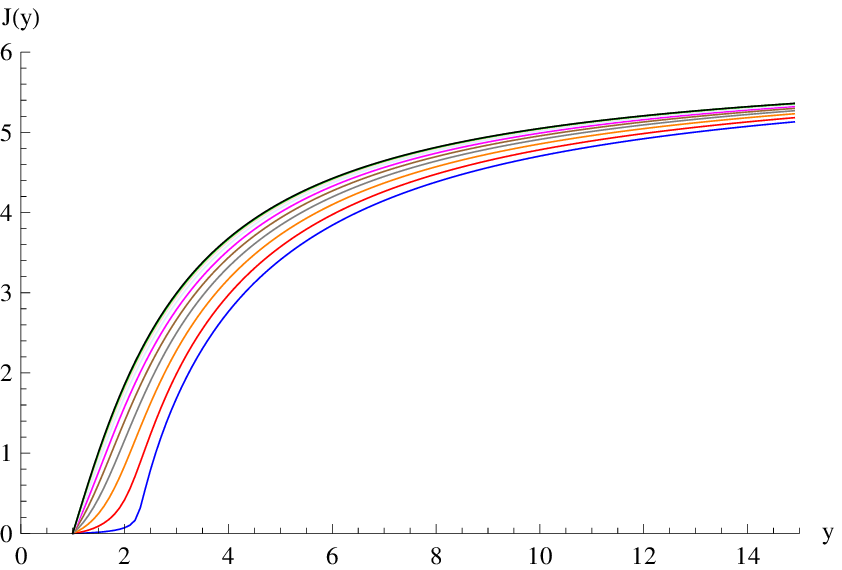}
\includegraphics[height=5cm,width=7.5cm]{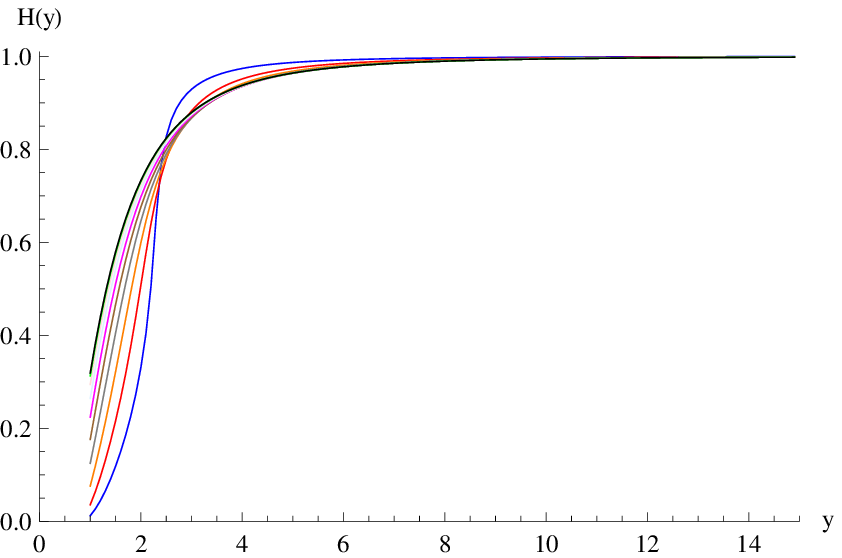}
\caption{
Solutions for the fields ($f(y), A(y), c(y), K(y), J(y), H(y)$) with the b.c. (\ref{bchor})-(\ref{bcinfty}) in the anisotropic case.
The curves correspond to $J_0=6$, $\hat m_{W}=0.4$ with different fixed values of
$\alpha=0.0$ (black), $0.1$ (green), $0.2$ (light red), $0.3$ (light blue), $0.4$ (fuchsia), $0.5$ (brown), $0.6$ (gray), $0.7$ (orange), $0.8$ (red), $0.88$ (blue).
One can appreciate from the curves corresponding to $f(y)/y^2$ the formation of a second  horizon at
$\alpha_c\approx 0.8825$.
The analogous solutions for the isotropic case can be found in \cite{LMS2}.
}
\end{center}
\end{figure}

\begin{figure}[H]\label{transicionIso}
\begin{center}
\includegraphics[height=6cm,width=8cm]{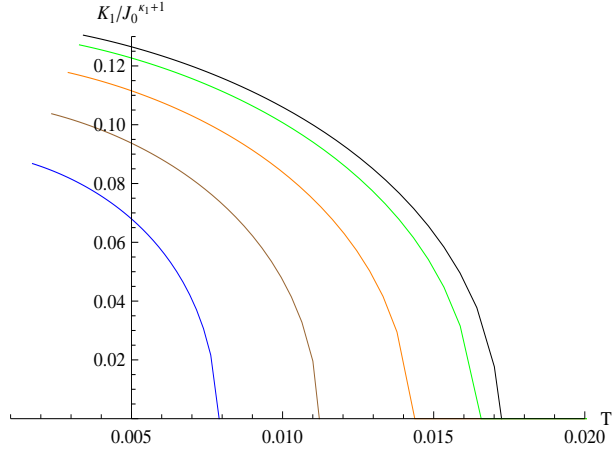}
\caption{
The order parameter $K_1/J_0{}^{1+\kappa_1}=\langle\hat{O}_{K_1}\rangle/J_0{}^{1+\kappa_1}$ is plotted
in the isotropic case at fixed $\alpha=0.7$, for different values of
$\hat m_{W}=0.01$ (black), $0.1$ (green), $0.2$ (orange), $0.3$ (brown), $0.4$ (blue) that correspond to the critical temperatures $T_c =0.017056, 0.016371, 0.014174, 0.011215, 0.00791$ respectively.
Near the critical temperature the order parameter behaves like $(T_{c}-T)^{1/2}$.
}
\end{center}
\end{figure}

\begin{figure}[H]\label{b}
\begin{center}
\includegraphics[height=6cm,width=8cm]{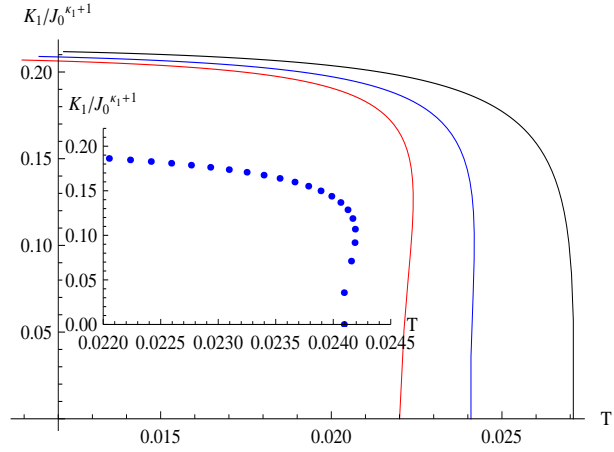}
\caption{
The order parameter $K_1/J_0{}^{1+\kappa_1}=\langle\hat{O}_{K_1}\rangle/J_0{}^{1+\kappa_1}$ is plotted
in the anisotropic case at fixed $\hat m_W=0.4$, for different values of
$\alpha=0.50$ (black), $0.53$ (blue), $0.55$ (red).
In the inset is displayed the multi-valuation of the order parameter for $\alpha_{tc} \approx 0.53$.
}
\end{center}
\end{figure}

\begin{figure}[H]\label{transicionAni}
\begin{center}
\includegraphics[height=6cm,width=8cm]{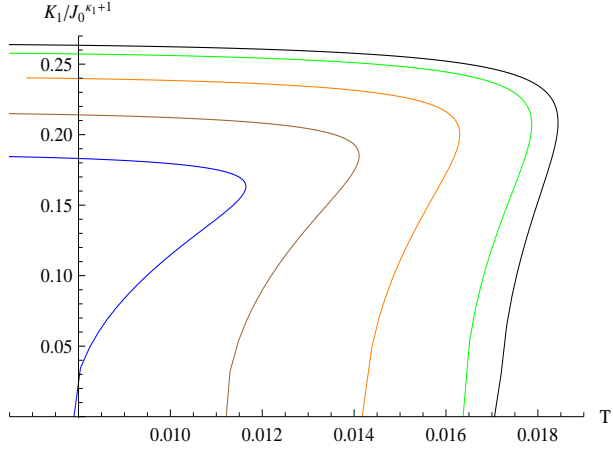}
\caption{
The order parameter  $K_1/J_0{}^{1+\kappa_1}=\langle\hat{O}_{K_1}\rangle/J_0{}^{1+\kappa_1}$ is plotted in
the anisotropic case at fixed $\alpha=0.7$, showing the phase transition at different values of $\hat m_{W}=0.01$ (black), $0.1$ (green), $0.2$ (orange), $0.3$ (brown), $0.4$ (blue), $0.32$ (blue).
}
\end{center}
\end{figure}

\subsection{The free energy}

According to the AdS/CFT correspondence, the free energy of the QFT is given by,
\be
F \equiv T\,S_{eucl}= \int^{\infty}_{y_h}dy \int d\vec x^2 L_{eucl}
\ee
From (\ref{S}) and using the e.o.m. the bulk contribution to the free energy density can be written as,
\bea\label{fbulk1}
f^{(bulk)} &=& \frac{1}{2\,e^2\,L^3}\,\int^{y_\infty}_{1}\,dy\,A(y)\,c(y)\,y^2\,
\left(\frac{6}{\alpha{}^2} -\frac{\lambda_0}{2}\,\left(H(y)^2-1\right)^2\right.\cr
&+&\frac{f(y)}{y^2}\,\left(\frac{K'_1(y)^2}{c(y)^2}+ K'_2(y)^2\right)
+\,\frac{K_1(y)^2\,K_2(y)^2}{y^4\,c(y)^2}\cr
&-&\left.\frac{J'(y)^2}{A(y)^2}-\frac{J(y)^2}{y^2\,f(y)\,A(y)^2}\,\left(\frac{K_1(y)^2}{c(y)^2} + K_2(y)^2\right)
\right)
\eea
The Gibbons-Hawking contribution is,
\be
f^{(GH)} = \frac{1}{2\,e^2\,L^3}\,\left(-\frac{2}{\alpha{}^2}\right)\,f(y)^\frac{1}{2}\,\left(
y^2\,f(y)^\frac{1}{2}\,A(y)\,c(y)\right)'|_{y_\infty}
\ee
Here we have introduced $y_\infty$ to regularize the expressions since they present divergent terms.
To this end we introduce a counter-term action \cite{HenninSken} \cite{BalaKraus},
\be
S^{(ct)}=\frac{1}{2\,\kappa^2}\;\int_{\partial{\cal M}} d^3x\,\sqrt{|h|}\; \frac{-2}{L}
\ee
which give rise to the following contribution to the free energy density,
\be
f^{(ct)} = \frac{1}{2\,e^2\,L^3}\,\frac{4}{\alpha^{2}} \left( f(y)^{1/2}A(y)c(y)y^{2}\right)|_{y_\infty}
\ee
The total free energy density of the system $f$ is then given by,
\be
f\equiv \lim_{y_\infty\rightarrow\infty}\left(f^{(bulk)}+f^{(GH)}+f^{(ct)}\right)
\ee
We remark that in order to analyze the results the right thing to do is to work with the scale invariant
free energy density,
\be\label{fsi}
\hat f \equiv \frac{\kappa^2}{L^2\,\mu^3}\; f
\ee

Figures $6$ and $7$ show the evolution of the free energy density (\ref{fsi}) with the mass
of the gauge boson for two different values of $\alpha$, in the isotropic and anisotropic cases respectively.
Figure $6$ displays the continuity of $\hat f$ at the critical  temperature (where the free energy density of the uncondensed phase intersects the curve of the condensed phase) for both values of $\alpha$, for any $\hat m_W$,
fact that indicates the second order character of the phase transition as the behavior of $K_1$ in figure $3$ suggested.
In figure $7$ instead it is observed the discontinuity in the first derivative of the free energy density at the critical temperature for $\alpha=0.7>\alpha_{tc}$ for any $\hat m_W$, signaling a first order phase transition.
In both cases the critical temperature decreases with growing $\hat m_W$, in agreement with the analysis of the
behavior of the order parameter made above.

In figures $8$ and $9$ the free energy densities of the isotropic and anisotropic phases are compared
for two values of the gravitational coupling, $\alpha<\alpha_{tc}$ (figure $8$) and $\alpha >\alpha_{tc}$
(figure $9$).
From them one can see that the free energy density of the anisotropic phase, no matter the region where the value of $\alpha$ is, i.e. if first or second order phase transitions take place, is lower than the
free energy density of the isotropic phase.
That is, the anisotropic phase is always energetically favored over the isotropic one.

\begin{figure}[H]
\begin{center}
\includegraphics[height=6cm,width=6cm]{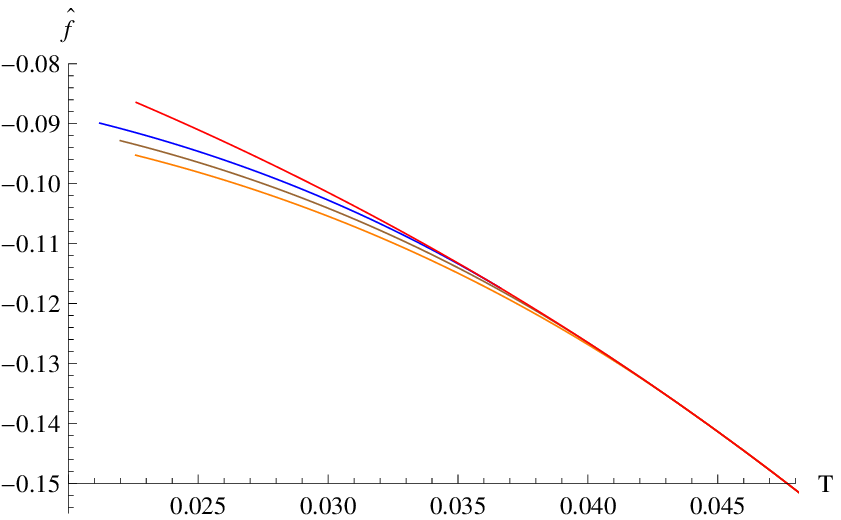}
\includegraphics[height=6cm,width=6cm]{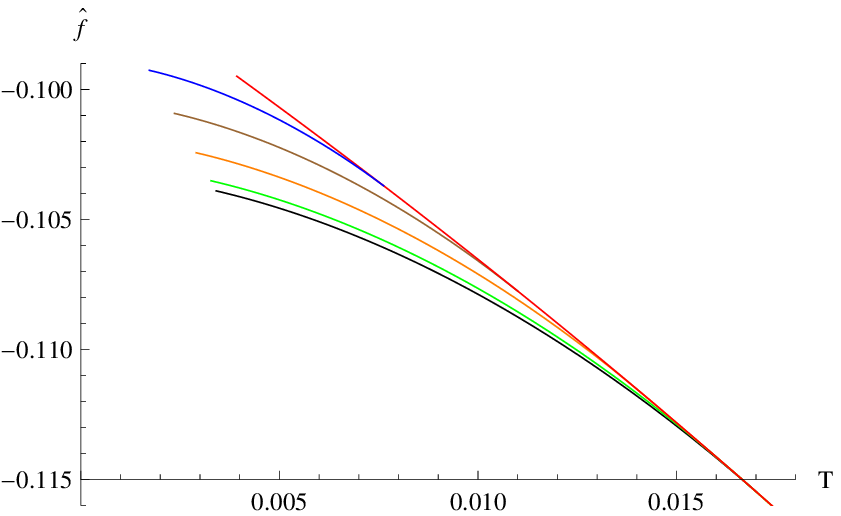}
\caption{
The free energy density $\hat{f}$ is plotted as function of the temperature in the isotropic case for two values of
the gravitational coupling constant, $\alpha =0.4$ (left) and $\alpha=0.7$ (right), at different values of
$\hat m_W =0.01$ (black), $0.1$ (green), $0.2$ (orange), $0.3$ (brown), $0.4$ (blue).
The red curve represents the free energy density of the uncondensed phase.
}
\end{center}
\end{figure}

\begin{figure}[H]
\begin{center}
\includegraphics[height=6cm,width=6cm]{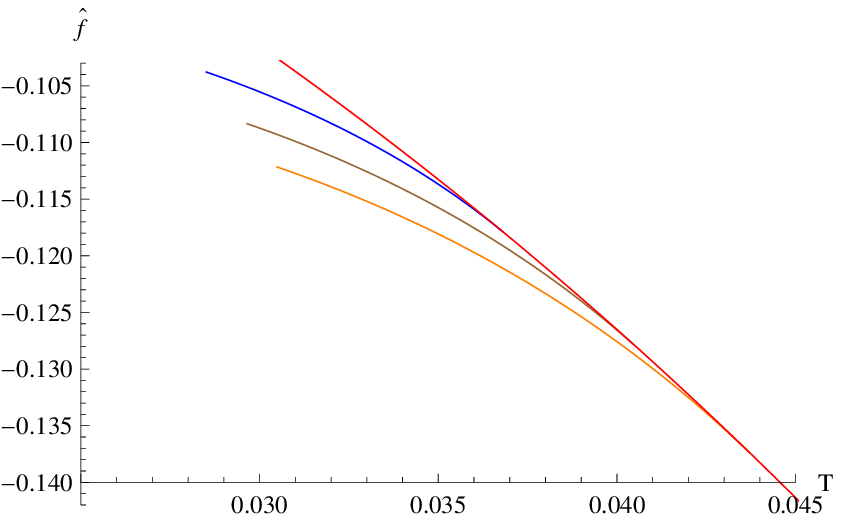}
\includegraphics[height=6cm,width=6cm]{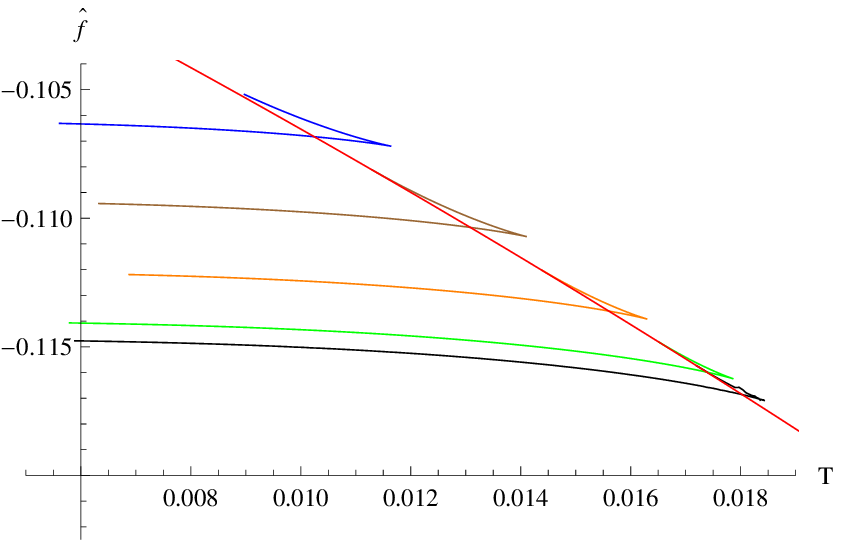}
\caption{
The free energy density $\hat{f}$ is plotted as function of the temperature in the anisotropic case for two values of
the gravitational coupling constant, $\alpha=0.4$ (left) and $\alpha=0.7$ (right), at different values of
$\hat m_W =0.01$ (black), $0.1$ (green), $0.2$ (orange), $0.3$ (brown), $0.4$ (blue).
The red curve represents the free energy density of the uncondensed phase.
}
\end{center}
\end{figure}

\begin{figure}[H]
\begin{center}
\includegraphics[height=5cm,width=5cm]{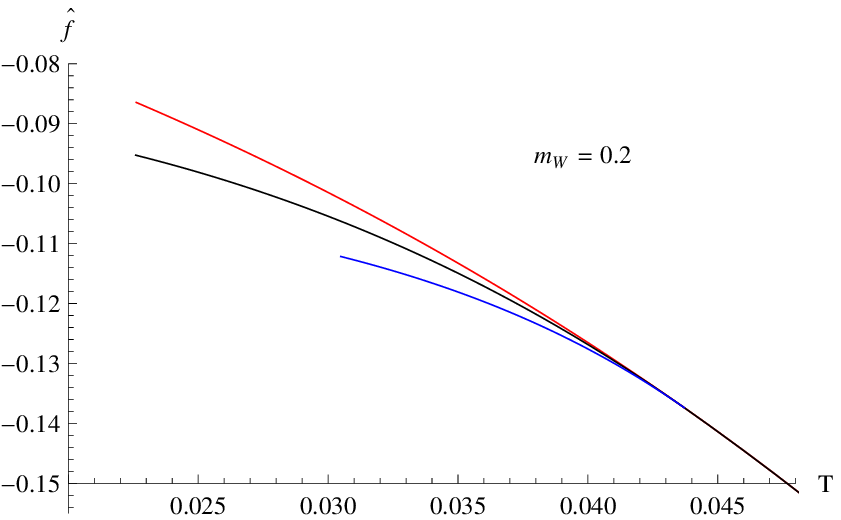}
\includegraphics[height=5cm,width=5cm]{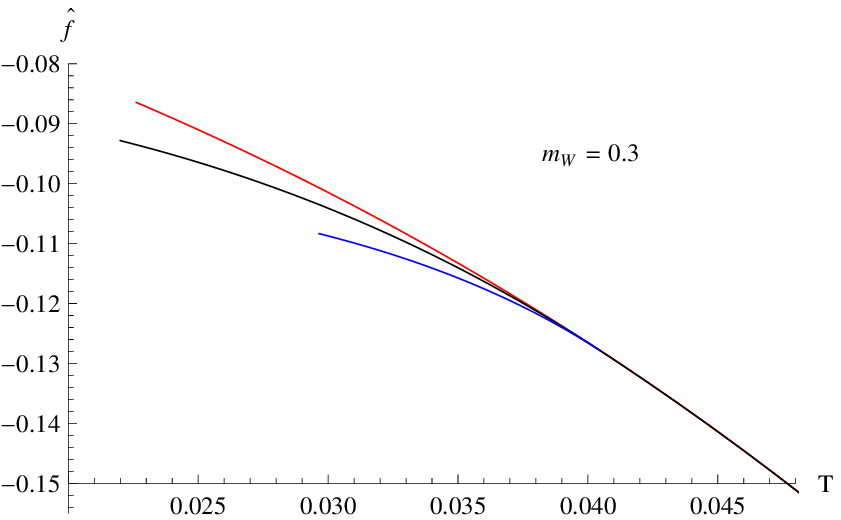}
\includegraphics[height=5cm,width=5cm]{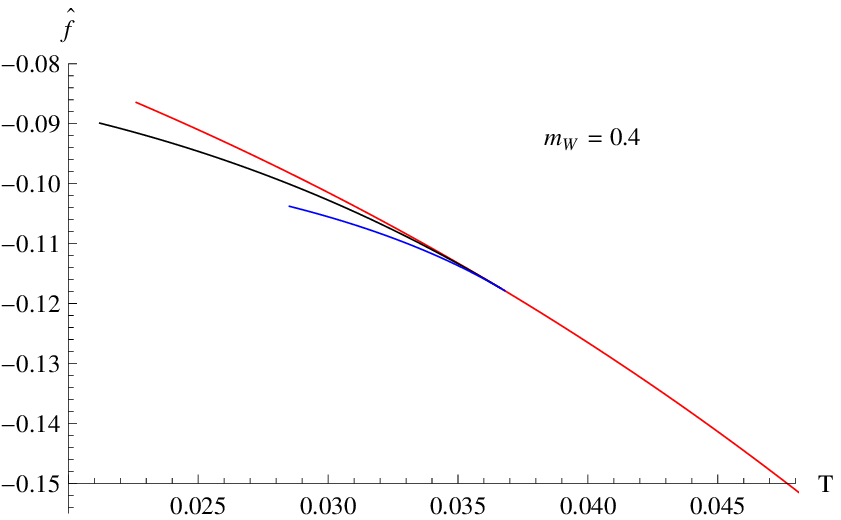}
\caption{
The free energy densities for the isotropic (black) and anisotropic (blue) cases are plotted as function of the temperature for different values of $\hat m_W$, at fixed $\alpha=0.4$.
The red curve represents the free energy density of the uncondensed phase.
}
\end{center}
\end{figure}

\begin{figure}[H]
\begin{center}
\includegraphics[height=5cm,width=5cm]{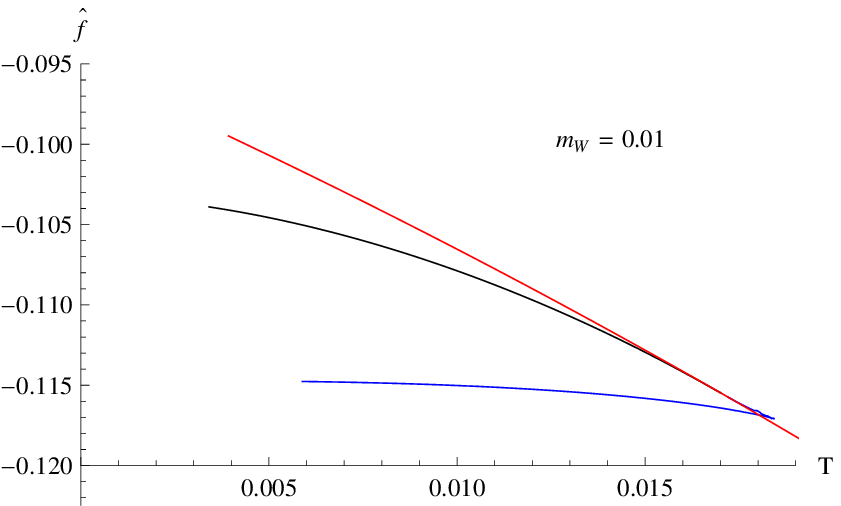}
\includegraphics[height=5cm,width=5cm]{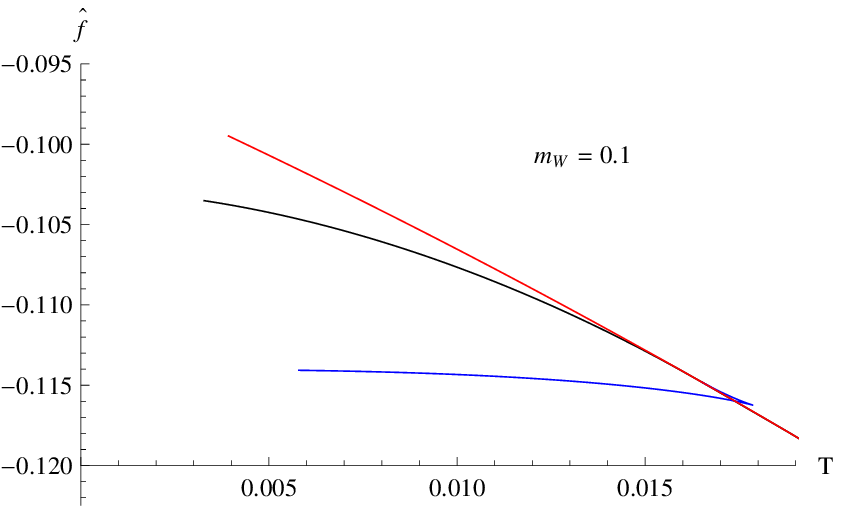}
\includegraphics[height=5cm,width=5cm]{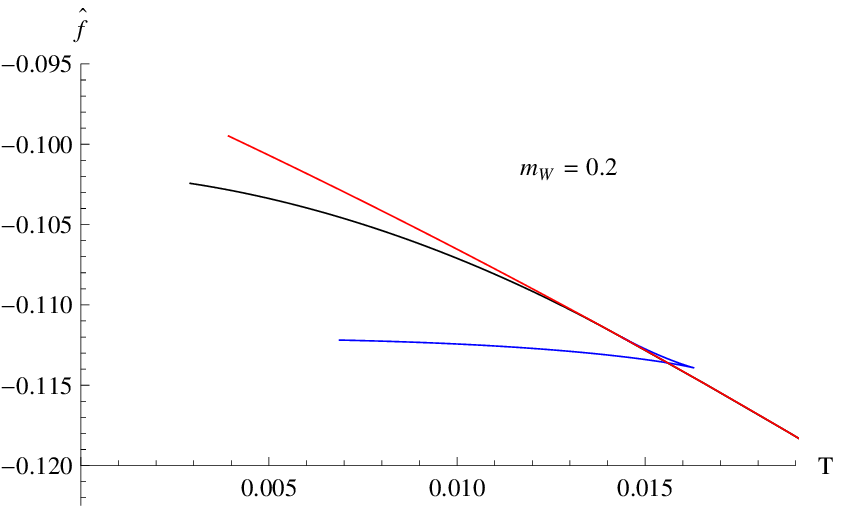}
\includegraphics[height=5cm,width=5cm]{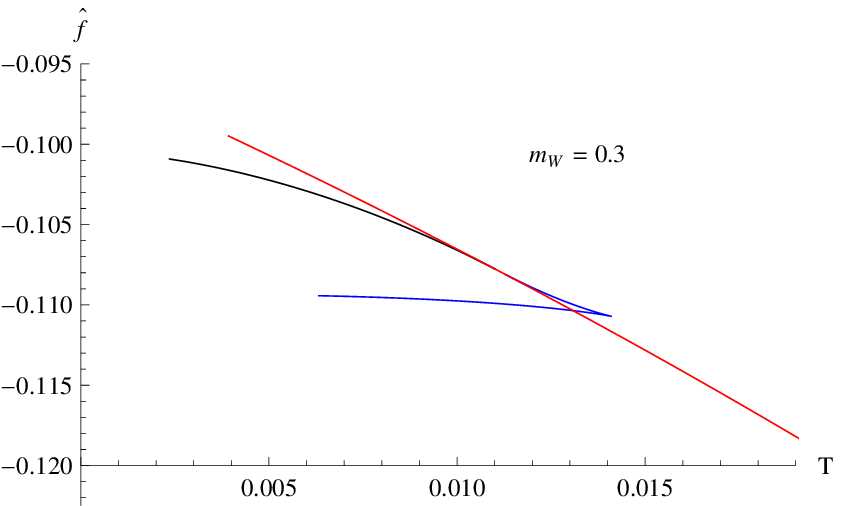}
\includegraphics[height=5cm,width=5cm]{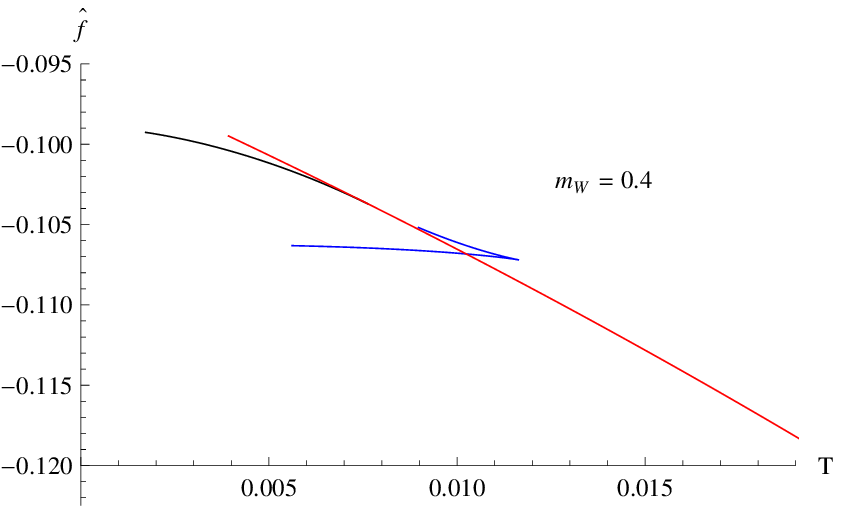}
\caption{
The free energy densities for the isotropic (black) and anisotropic (blue) cases are plotted as function of the temperature for different values of $\hat m_W$, at fixed $\alpha=0.7$.
The red curve represents the free energy density of the uncondensed phase.
}
\end{center}
\end{figure}

\section{Zero temperature solutions
\label{T=0}
}
\cleqn

In this section we will address the problem of quantum phase transitions in three dimensional $p$-wave superconductors (anisotropic case),
i.e. transitions at $T=0$, that to our knowledge was not considered before in the literature (see however \cite{Alishahiha} \cite{Zerolimit} \cite{Zerolimit2} \cite{Zerolimit3} for related studies in other settings).

It is known that when a charged AdS black hole is driven to a state of zero temperature it becomes extremal,
but its entropy is different from zero and then it can not describe the ground state of the superconductor that
we are presumably modeling holographically.
To reach our goal the radius of the black hole needs to become null, to agree with the third law of thermodynamics and really describe the quantum ground state \cite{Basu:2009vv} \cite{basesuperconductor}.
So we must impose that $y_h=0$, i.e. the coordinate $y\in[0,\infty)$.
A very important thing from a technical point of view is that while the asymptotic behavior of the fields is as in (\ref{bcinfty}), the expansions near the horizon drastically change with respect to the $T>0$ case.
At leading order the (non analytical) behavior of the fields for $y\rightarrow 0^+$ is,
\bea\label{bchorT=0}
f(y)&=&y^{2}- \alpha{}^2\;
\frac{\tilde{k}_0\,\tilde{j}_0{}^{2}}{2\,\tilde{c}_0\,\tilde{a}_0{}^{2}}\,
\frac{e^{-\frac{2\tilde{k}_{0}}{\tilde{c}_{0} y}}}{y} +\dots\cr
A(y)&=&\tilde{a}_0\,\left(1+\alpha{}^2\,\frac{\tilde{k}_0{}\,\tilde{j}_0{}^2}{\tilde{c}_0{}\tilde{a}_0{}^2}\,
\frac{ e^{-\frac{2\tilde{k}_{0}}{\tilde{c}_{0} y}} }{y^3}+\dots\right)\cr
c(y)&=&\tilde{c}_{0}\,\left( 1 + \alpha{}^2\;
\frac{\tilde j_0{}^2}{\tilde a_0{}^{2}}\;\frac{e^{-\frac{2\tilde{k}_{0}}{\tilde{c}_{0} y}}}{y^2}+\dots\right)\cr
K(y)&=&\tilde{k}_{0}\,\left(1 - \frac{\tilde c_0{}^2\,\tilde j_0{}^2}
{4\tilde a_0{}^2\,\tilde k _0{}^2}\;e^{-\frac{2\tilde{k}_{0}}{\tilde{c}_{0} y}}+\dots\right)\cr
J(y)&=&\tilde j_0\;e^{-\frac{\tilde{k}_{0}}{\tilde{c}_{0} y}}+\dots\cr
H(y)&=&\tilde h_0\;e^{-\frac{\tilde{k}_{0}}{\tilde{c}_{0} y}}+\dots
\eea
The independent constants are $\tilde{k}_{0},\tilde{j}_{0},\tilde{h}_{0},\tilde{a}_{0},\tilde{c}_{0}$.
Such constants are chosen in the same way as in the $T>0$ case, see (\ref{bcdefinitivas}).
From (\ref{bchorT=0}) it follows that near the horizon the solution is another $AdS_4$ space.
The solutions that describe the quantum ground state of the superconductor in the condensed phase
are therefore domain walls interpolating $AdS_4$ spaces with the same radius $L$ but different light velocities in both
directions, in virtue of the fact that $\tilde a_0\neq 1$ and $\tilde c_0 \neq 1$.
More explicitly,
\be
\frac{v_1^{UV}}{v_1^{IR}} = \frac{y_h}{y_\infty}\,\frac{c(y_h)}{c(y_\infty)}\,
\sqrt{\frac{f(y_\infty)}{f(y_h)}}\,\frac{A(y_\infty)}{A(y_h)}= \frac{\tilde{c}_{0}}{\tilde{a}_{0}}\quad;\quad
\frac{v_2^{UV}}{v_2^{IR}} = \frac{y_h}{y_\infty}\,
\sqrt{\frac{f(y_\infty)}{f(y_h)}}\,\frac{A(y_\infty)}{A(y_h)}= \frac{1}{\tilde{a}_{0}}
\ee
On the other hand, the uncondensed phase is described strictly by $AdS_4$ space and $J(y)=J_0$,
which replace the AdS-RN solution (\ref{uncond1})-(\ref{uncond2}).
Interestingly, we found that above a certain $\tilde\alpha$ the solution representing the condensed phase
disappears and the only solution that exists is AdS space.
This result can be guessed from the following analysis borrowed from \cite{Zerolimit} (see also \cite{Basu:2009vv}).
At very low temperatures the normal phase is nearly represented by the extremal, zero temperature  Reissner-N\"ordstrom solution (\ref{uncond1}-\ref{uncond3}) with $J_0{}^2=\frac{6}{\alpha^2}$,
whose near horizon geometry is $AdS_2\times\Re^2$.
If we perturb this solution with a non-zero gauge field $K_1(y)=K(y)$, from the first equation in (\ref{matteom1})
its linear equation in this background results,
\be
0=\left(\rho^2\,\partial_\rho^2 + 2\,\rho\,\partial_\rho - \hat m_{eff}{}^2\right)\,K(y)
\ee
where $\rho\equiv y-1$, which is just the wave equation for $AdS_2$ with an effective mass,
\be
\hat m_{eff}{}^2 = \frac{1}{6}\,\left(\hat m_W{}^2 - \frac{1}{\alpha^2}\right)
\ee
So, the instability to form $SU(2)$ vector hair at low temperature is just that of scalar fields below
the BF bound for $AdS_2$, $\hat m_{BF}^2 = -\frac{1}{4}$.
That is, when $\hat m_{eff}{}^2<\hat m_{BF}^2$ one could wait that AdS vacuum gets unstable and
the system prefers to be in the phase described by the superconducting black hole solution with non abelian hair.
Thus we get a plausible condition for instability,
\be\label{alfaguess}
\alpha^2 <\tilde\alpha_{guess}^2 \equiv \frac{1}{\frac{3}{2}+\hat m_W{}^2}
\ee

Figure $10$ shows the fields for different values of the gravitational coupling
\footnote{We use the scaling symmetry (\ref{si2}) to fix $J_0 =1$.
}.
In the example showed ($\hat m_W=0.4$) we obtain $\tilde\alpha\approx 0.8825 \approx\alpha_c|_{T>0}$,
see figures $2$ and $10$.

\begin{figure}[H]\label{T0Solution}
\begin{center}
\includegraphics[height=3.2cm,width=3.9cm]{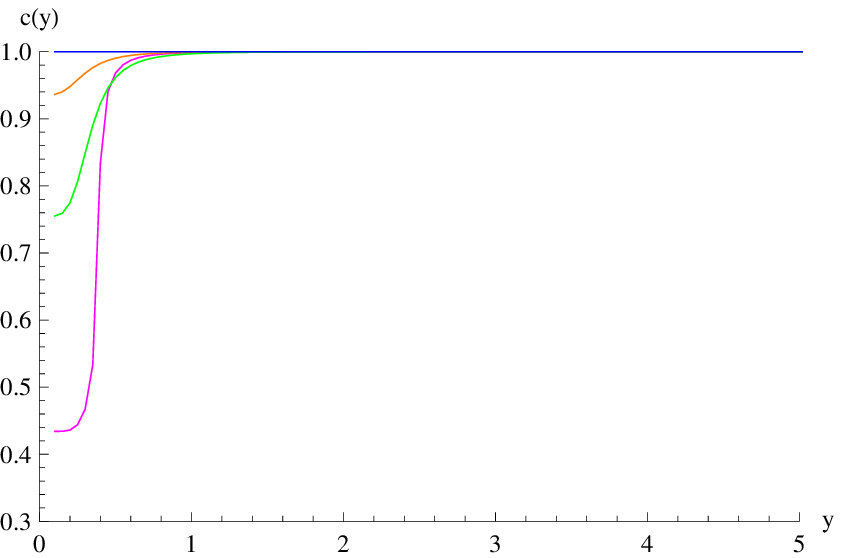}
\includegraphics[height=3.2cm,width=3.9cm]{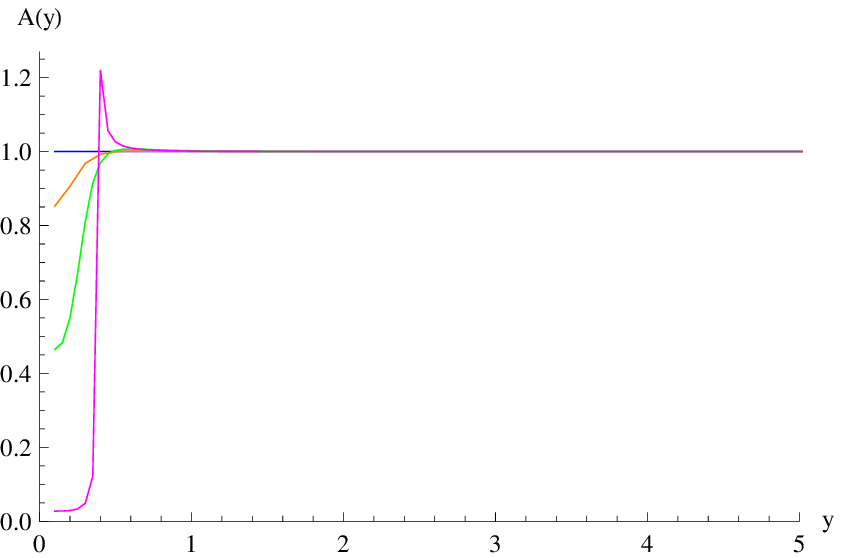}
\includegraphics[height=3.2cm,width=3.9cm]{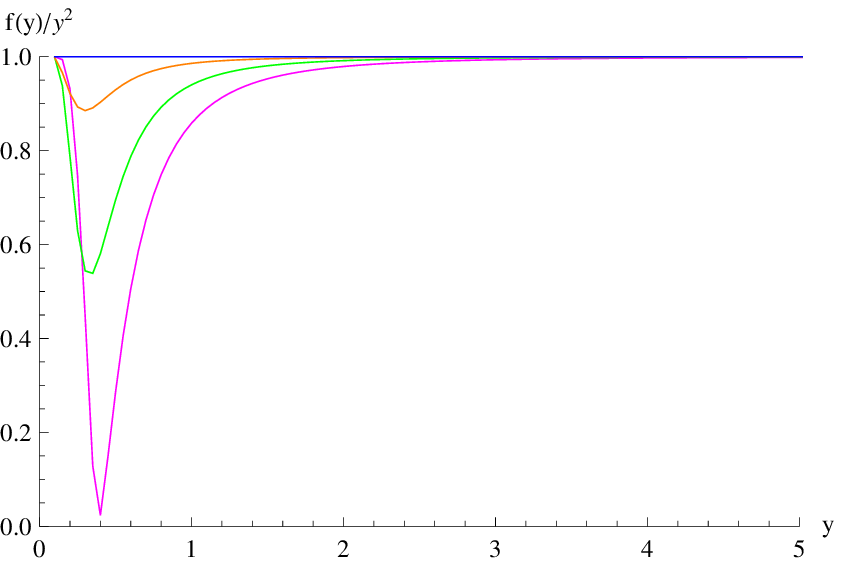}\\
\includegraphics[height=3.2cm,width=3.9cm]{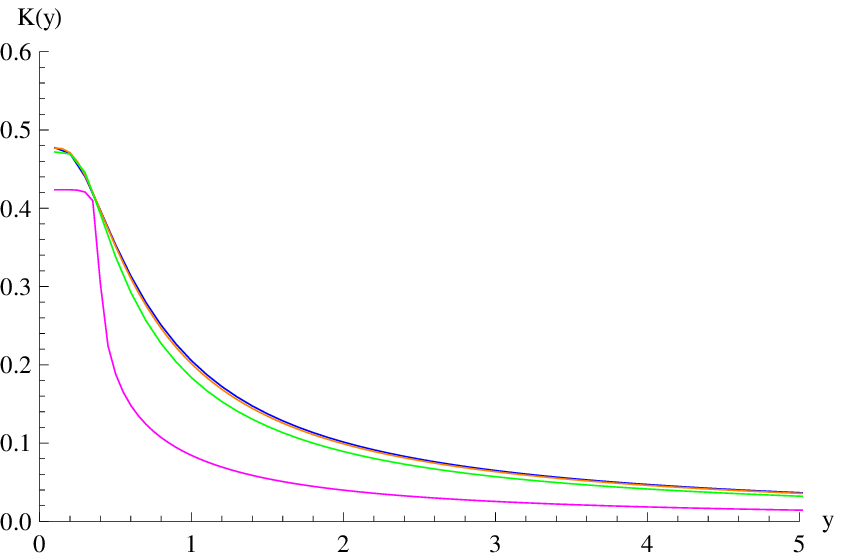}
\includegraphics[height=3.2cm,width=3.9cm]{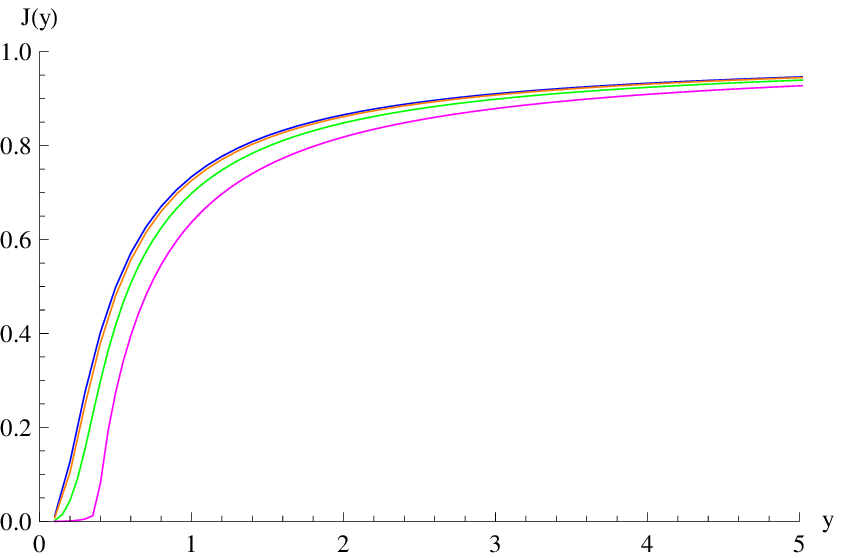}
\includegraphics[height=3.2cm,width=3.9cm]{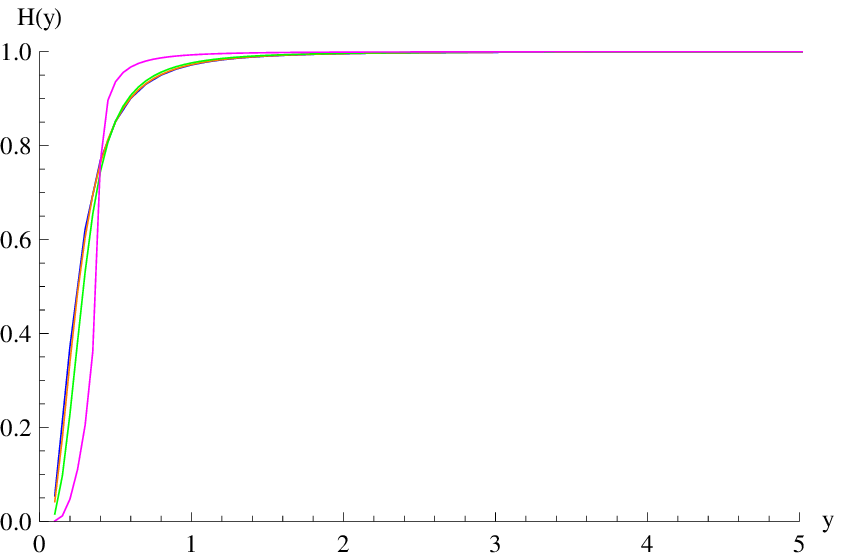}
\caption{We display for $T=0$ the solutions
$f(y)/y^{2}$, $A(y)$, $c(y)$, $H(y)$, $K(y)$ and $J(y)$
as functions of the coordinate $y$ from the horizon, now located at $y=0$,
towards the boundary, for $\hat m_W=0.4$ and different values of
$\alpha=0.0$ (blue), $0.3$ (orange), $0.6$ (green), $0.8$ (fuchsia).
One can appreciate from the curves corresponding to $f(y)/y^2$ the formation of a second  horizon when  $\alpha$ approaches $\tilde\alpha\approx\alpha_c|_{T>0}\approx 0.8825$; for larger values of $\alpha$ the asymmetric solution ceases to exist.
}
\end{center}
\end{figure}

We worked out the solutions for different values of the parameter $\hat m_{W}$, in Table I the corresponding critical couplings are shown.
It is observed that $\tilde\alpha$ decreases with growing $\hat m_W$ and that $\tilde\alpha_{guess}<\tilde\alpha$, what is consistent
with the instability analysis made before.

On the other hand, in Figure $11$ it is shown the order parameter as a function of the coupling $\alpha$.
We have verified that the behavior near the critical coupling is of the type,
\be\label{K1alfa}
K_1(\alpha) \sim \left( \tilde\alpha - \alpha\right)^\frac{1}{2}
\ee
consistent with the existence of a second order phase transition in the mean field, large number of degrees of freedom limit.

\begin{figure}[H]\label{T0Solution2}
\begin{center}
\includegraphics[height=5cm,width=5cm]{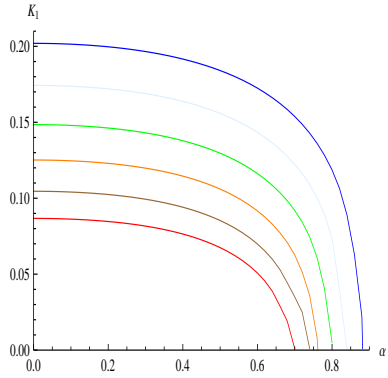}
\caption{The order parameter $K_{1}$ is plotted as function of $\alpha$ at $T=0$, for different values of $\hat m_W=0.5$ (blue), $0.6$ (lightblue),
$0.7$ (green), $0.8$ (orange), $0.9$ (brown), $1.0$ (red).
The values for which $K_{1}=0$ define the critical couplings $\tilde\alpha$. }
\end{center}
\end{figure}

\begin{table}
\begin{center}\label{TABLE}
\begin{tabular}{|c|c|c|c||c||c||c||c|}
\hline
$\hat m_{W}$ & $\tilde\alpha$  \tabularnewline
\hline
$0.1$ & 1.03   \tabularnewline
\hline
$0.2$  & 0.97   \tabularnewline
\hline
$0.3$  & 0.94  \tabularnewline
\hline
$0.4$  & 0.88   \tabularnewline
\hline
$0.5$ & 0.86    \tabularnewline
\hline
$0.6$ & 0.84   \tabularnewline
\hline
$0.7$ & 0.80  \tabularnewline
\hline
$0.8$ & 0.76 \tabularnewline
\hline
$0.9$ & 0.74  \tabularnewline
\hline
$1.0$ & 0.70  \tabularnewline
\hline
\end{tabular}
\caption{Critical gravitational couplings $\tilde\alpha$ for different values of $\hat m_W$ at $T=0$.}
\end{center}
\end{table}

\section{Analysis for $\lambda\neq 0$}

In this section we will study the effect of a non-zero Higgs potential as introduced in (\ref{S}),
specified by the Higgs vev scale $H_0$ and the strength $\lambda$.
For simplicity we will work in the no back-reaction limit $\alpha=0$, although the new insights
does not depend on this fact.

In the conventions of Appendix B, the e.o.m. (\ref{matteom1}) in the anisotropic case reduce to,
\bea\label{eom_lambda}
\left(f(u)\,K^{\prime}(u)\right)^{\prime}&=& \left( \hat m_W{}^2\,\frac{H(u)^{2}}{u^{2}}-\frac{J(u)^{2}}{f(u)}\right)\,K(u)\cr
u^{2}\,\left(\frac{f(u)}{u^{2}}\,H^{\prime}(u)\right)^{\prime}&=&\left( K(u)^{2}+
\frac{\lambda_0}{\hat m_W{}^2}\,\frac{H(u)^{2}-1}{u^{2}}\right)\,H(u)\cr
J''(u)&=&\frac{K(u)^{2}}{f(u)}\,J(u)
\eea
where $f(u)=1-u^{3}$.

The existence of a non-zero $\lambda$ does not modify the behavior of $K(y)$ and $J(y)$ on the boundary that
remain as in (\ref{bcinfty}), but it does in the Higgs case where we now have,
\begin{equation}
H(u)=1 + H_{-}\,u^{\Delta_-}+\dots+H_{+}\,u^{\Delta_+}+\dots
\end{equation}
where, for general $\lambda_0$ and $\hat m_W$,
\be
\Delta_\pm = \frac{3}{2} \pm\sqrt{\frac{9}{4} + 2\,\frac{\lambda_0}{\hat m_W{}^2}}
\ee
A first well-known fact is that reality of $\Delta_\pm$ necessarily implies the BF bound $\lambda_0\geq-\frac{9}{8}\,\hat m_W{}^2$.
When we are in the window  $-\frac{9}{8}\leq\frac{\lambda_0}{\hat m_W{}^2}\leq-\frac{3}{4}$ both modes are normalizable and lead to
consistent quantization and we can impose $H_{-}=0$ or $H_{+}=0$.
If $\lambda_{0}>-\frac{3}{4}\,\hat m_W{}^2$, the condition $H_{-}=0$  must be imposed \cite{Bolognesi:2010nb}.
We will consider for definiteness the case $\lambda_{0}> 0$.

A very interesting fact is that, besides the existence of a bound from below for the Higgs coupling as stated above,
a straight analysis of the solution near the boundary $u=0$ yields the result that a bound from above is also present.
We find that there exists a critical value $\lambda_0^c$ defined by,
\be\label{lambdacritico}
\frac{\lambda_0^c}{\hat m_W{}^2} = 2 + (\kappa_1 - 1)\,\left(2\,\kappa_1 +3\right)
\ee
such that for $\lambda_0>\lambda_0^c$ the condensed solution ceases to exist.
In the example considered below $\hat m_W{}^2 =1$, $\lambda_0^c\sim 5.854$.
This is so for both the isotropic and anisotropic cases.

\begin{figure}[H]\label{lambdavariable1}
\begin{center}
\includegraphics[height=5cm,width=5cm]{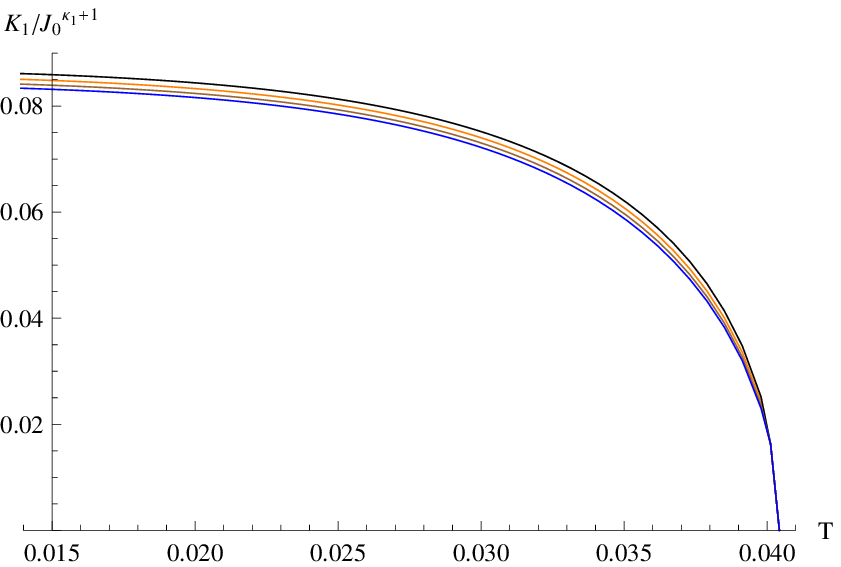}
\includegraphics[height=5cm,width=5cm]{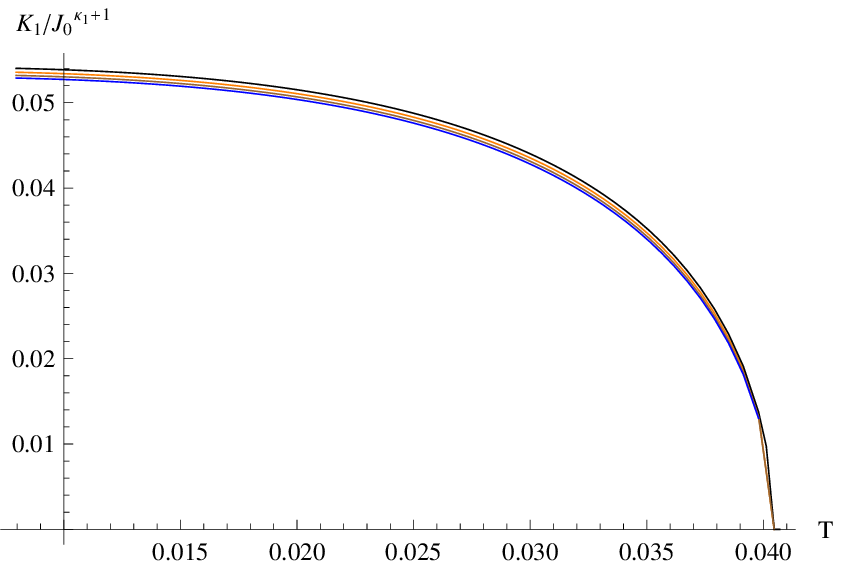}
\caption{
The condensate $K_{1}$ as a function of the temperature $T$, for $m_W{}^2=1$ and
$\lambda_{0}=0$ (black), $0.25$ (orange), $0.5$ (brown),$0.75$ (blue), in the anisotropic (left) and isotropic (right) cases.
}
\end{center}
\end{figure}


\begin{figure}[H]\label{freeenerylambdavariable11}
\begin{center}
\includegraphics[height=5cm,width=5cm]{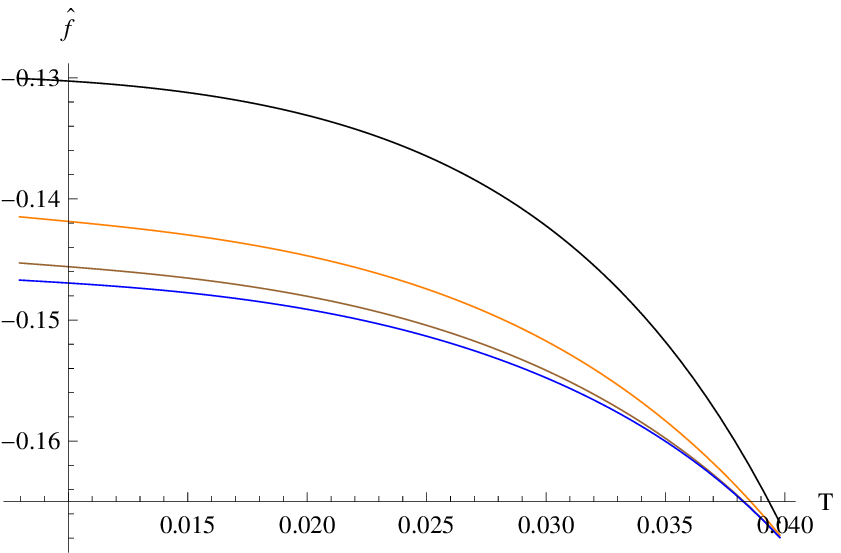}
\includegraphics[height=5cm,width=5cm]{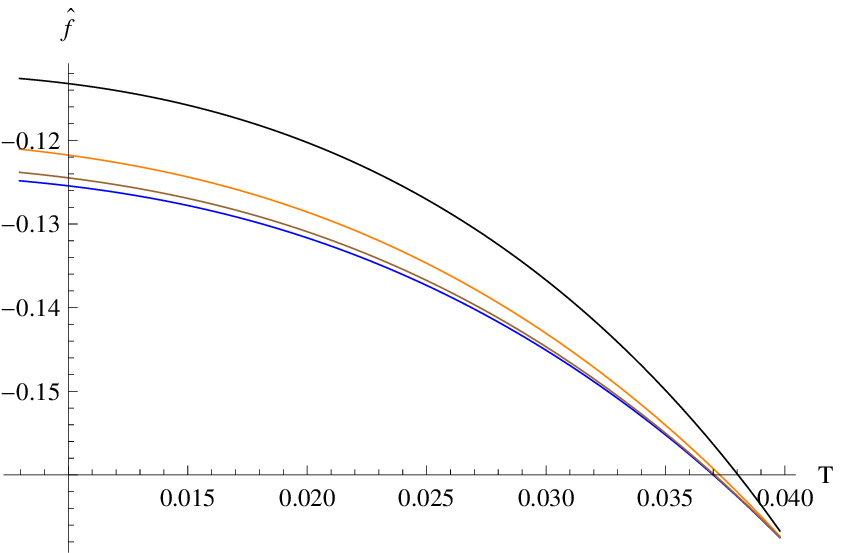}
\caption{
The free energy $\hat{f}$ as a function of the temperature $T$ for $m_W{}^2=1$ and
$\lambda_{0}=0$ (black), $0.25$ (orange), $0.5$ (brown), $0.75$ (blue) in the anisotropic (left) and isotropic (right) cases.
}
\end{center}
\end{figure}

\begin{figure}[H]\label{freeenerylambdavariable2}
\begin{center}
\includegraphics[height=5cm,width=4cm]{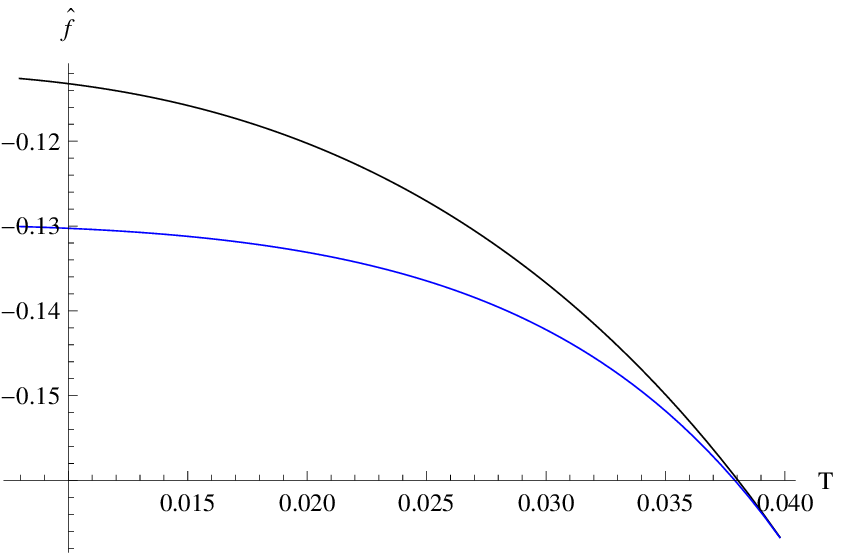}
\includegraphics[height=5cm,width=4cm]{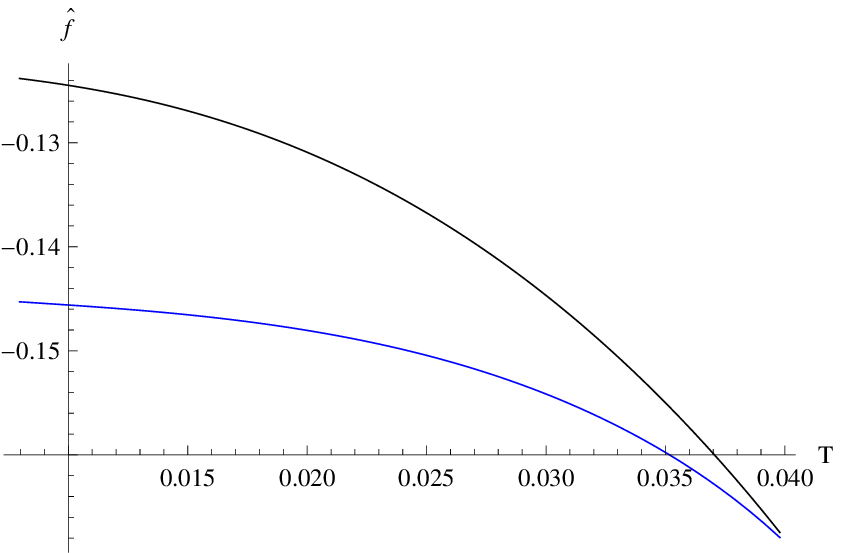}
\includegraphics[height=5cm,width=4cm]{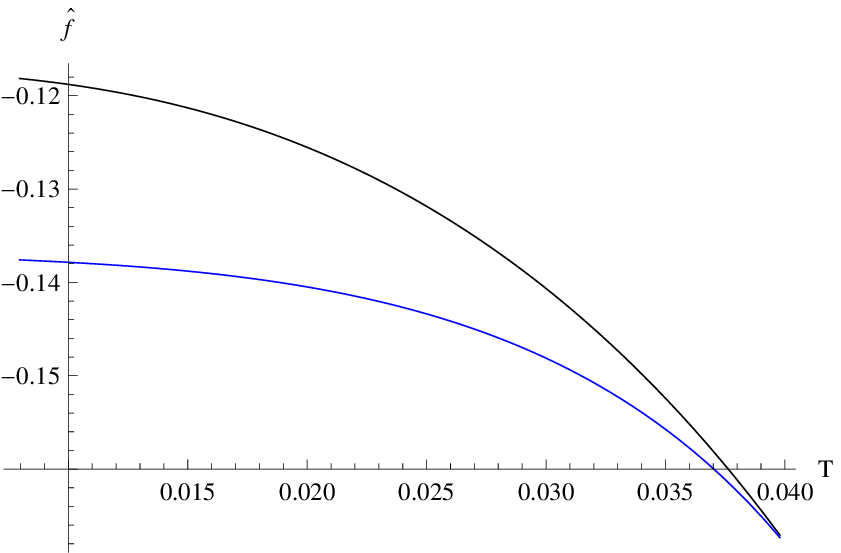}
\includegraphics[height=5cm,width=4cm]{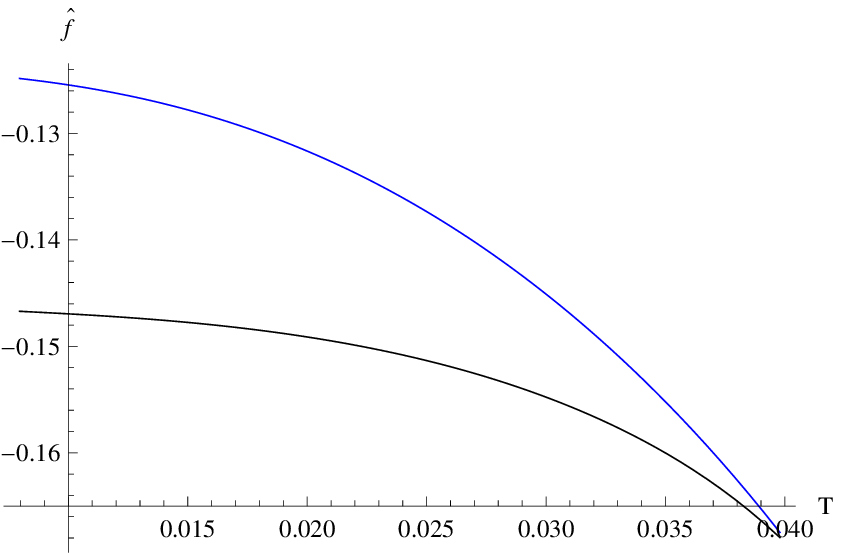}
\caption{Comparison between the free energies of the anisotropic (blue) and isotropic (black) cases for $m_W{}^2=1$ and, from left to right,
$\lambda_0=0, 0.25, 0.5 ,0.75 $.
}
\end{center}
\end{figure}

\begin{figure}[H]\label{freeenerylambdavariable3}
\begin{center}
\includegraphics[height=5cm,width=5cm]{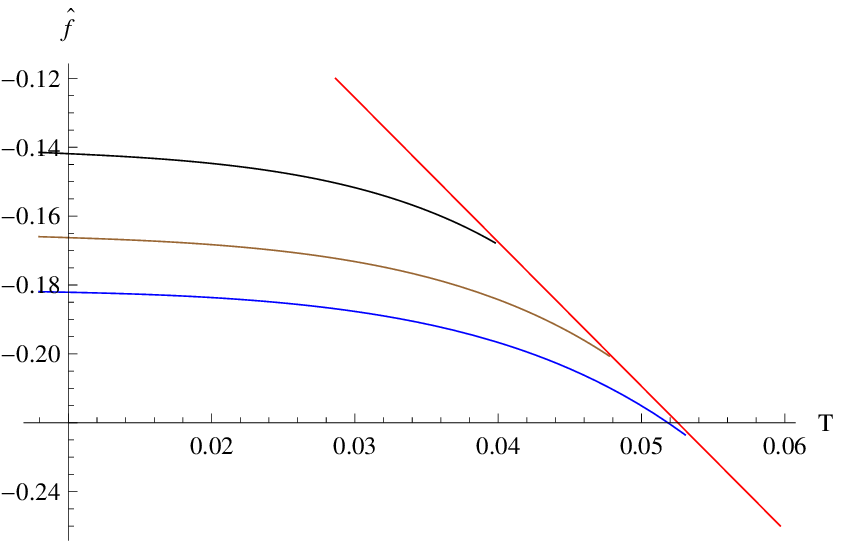}
\includegraphics[height=5cm,width=5cm]{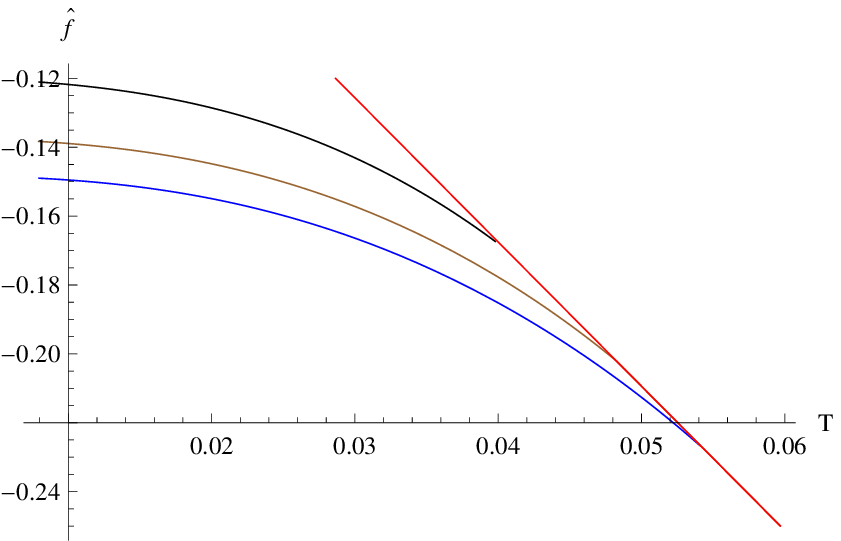}
\caption{The free energy as a function of the temperature in the anisotropic (left) and isotropic (right) cases,
for fixed $\lambda_0=0.25$ and different values of
$\hat m_W=0.5$ (blue), $0.7$ (brown) and $1.0$ (black).
The red curve represents the free energy of the normal phase.
}
\end{center}
\end{figure}

Figures $12$ and $13$ show the condensate and the free energy respectively as functions
of the temperature, for a fixed $\hat m_W$ and different Higgs couplings
\footnote{
For higher Higgs strengths towards the critical value the curves does not experiment significative changes.
}.
One can appreciate that both the order parameter and the free energy decreases with increasing strength of the potential $\lambda_{0}$; however the critical temperature does not change and the phase transitions remain of second order.

In Figure $14$ the free energies of both isotropic and anisotropic cases as functions of the temperature are plotted
together for comparison, for fixed Higgs scale and different Higgs couplings.
The anisotropic phase always remains energetically favored over the isotropic one.

As stated before, a critical value (\ref{lambdacritico}) above which the condensed phase does not exist is present,
and it was verified by our numerical calculations; such result occurs in both isotropic and anisotropic cases.

Figures $15$ show the free energies in function of the temperature for a fixed $\lambda_{0}$, for different
gauge boson masses.
It is observed that they increase with increasing $\hat m_W$.
The curves are similar to the left curves in figures $6$ and $7$.

\section{Conclusions and outlook}

In this paper we have investigated four dimensional solutions of black holes with non-abelian,
$SU(2)$ hair introduced by Yang-Mills gauge bosons and a non trivial Higgs field in the adjoint representation,
whose v.e.v. triggers the breaking of the gauge symmetry to a $U(1)$ subgroup under which the black hole is charged.

In the spirit of the AdS/CFT correspondence, the symmetric solution given by
the AdS-RN black hole when the temperature is positive and AdS when $T=0$,
describes the uncondensed phase of the dual three dimensional QFT.
A solution with non-abelian hair generically breaks the $U(1)$ gauge symmetry
together with the rotational symmetry, and is interpreted as describing a condensed phase of the QFT.
The order parameter is the coefficient of the leading order term  of
the magnetic component of the gauge field, and thus the systems described
are generically termed $p$-wave superfluids/superconductors.
We have considered two cases.
The isotropic case that describes $p+ip$-wave superconductors
where the diagonal subgroup of $U(1)_{gauge}\times SO(2)_{rot}$ is preserved,
and the anisotropic case where no symmetry is preserved.
In both cases we get phase transitions at critical temperatures that
decrease when the gravitational coupling grows and in the case of anisotropic superconductors the phase transitions become of first order for large gravitational couplings \cite{Ammon:2009xh}. 
These results are summarized in the phase diagrams presented in figure $1$. 

We also find solutions that describe the zero entropy ground state of the $p$-wave superconductor,
showing the existence of phase transitions from the normal phase (described by AdS space) to this condensed
phase, that is present below a certain value of the gravitational coupling $\tilde\alpha$.
These transitions are of second order, according to the behavior (\ref{K1alfa}) of the order parameter near the transition obtained from figure $11$.
Such states are described by domain wall geometries that interpolate two AdS spaces.
The occurrence of AdS space near the horizon, with the the same scale as the AdS in the boundary but different 
light velocities, presumably indicates that there is an emergent scale invariance in the $T=0$ limit
\cite{Ammon:2009xh} \cite{Basu:2008st}.

Finally we study the effect of considering a non zero Higgs potential.
It was found that for Higgs coupling constants greater than a critical value $\lambda_c$,
the solution collapses to the normal one.
This fact relies on the ultraviolet behavior of the system; in particular is independent if the back-reaction
is considered or not.
For $\lambda$ below $\lambda_c$ the system and its thermodynamic variables behave qualitatively as in the case $\lambda =0$, but with lower free energy.

A very relevant fact conjectured in the literature \cite{GubserNonAb} \cite{GubserP} \cite{Basu:2009vv} 
for systems without Higgs fields that we have explicitly addressed in this paper including them together with the corresponding Higgs potential, is that below the critical temperature in all the parameter space we found that the free energy density of the anisotropic solution is lower than that of the isotropic one, indicating that the $p$-wave superconductor phase is more stable that the one corresponding to the $p+ip$-wave superconductor. 
This result is illustrated in figures $8$, $9$  and $14$.

We believe it is worth to make the following remarks.
It is straight to see that if we switch off the Higgs field the e.o.m. of the $p$-wave superconductor are recovered.
However if we switch off the magnetic part of the gauge field, $K_{1}=K_{2}=0$, it is seen that we do not
recover the e.o.m. of a $s$-wave superconductor.
This is due to the fact that we are switching the Higgs field in the $X_{0}$-direction, not in the
$X_{1}$-direction.
This lead us to conclude that in our set-up the Higgs field can not condense spontaneously since the temporal
component of the gauge field (which plays a fundamental role in the condensation) is null.
Therefore we will never have competition between s-wave and p-wave phases, as it takes place in the cases analyzed 
in references \cite{Nie:2013sda}-\cite{Nie:2014qma} 
where the matter field ansatz is slightly different and the vev Higgs field is put to zero. 
Furthermore it is not difficult to see from the e.o.m. that a configuration where a vev (\ref{Jvev}) is present
necessary implies a non-trivial Higgs field; however the vev of its dual scalar operator ${\cal O}(x)$,
\be 
\langle{\cal O}(x)\rangle\sim H_1
\ee
does not indicate any spontaneous breaking in view of the presence of the source $H_0$.

We stress that, although they share some similar characteristics, the presence of the Higgs fields with the non trivial b.c. $|\vec H(\infty)|=H_0>0$, introduces a scale that  makes our systems different from those considered in precedence since \cite{GubserP} \cite{Alishahiha}, in which the scalars were not present.
On one hand from the obvious fact that the system is larger and more complex; in particular we have three free parameters ($\alpha, \hat m_W, J_{0}$) and, among other things, the dimension (\ref{dimOP}) of the order parameter remains arbitrary.
Instead, in EYM systems where the Higgs field is not present, the temperature for example is a function of just one parameter $\alpha\equiv\frac{\kappa}{e\,L}$ \cite{Ammon:2009xh}.
On the other hand and more important, from the QFT point of view to which the systems we have considered 
are presumed to be holographically dual.
This fact can be elucidate by studying the transport properties of the system, i.e. the conductivities.
Even ignoring back-reaction effects, we finish with a system of fifteen coupled second order equations that
results much more cumbersome to disentangle than in the Abelian case or in the absence of Higgs fields.
We hope to report results in this direction in a near future \cite{gl3}.

\section*{Acknowledgments}

We would like to thank Nicol\'as Grandi and Ignacio Salazar Landea for encouragement and continuous support,
and Borut Bajc, Jorge Russo and Gast\'on Giribet for careful reading of the manuscript and useful comments.
This work was supported in part by CONICET, Argentina.

\appendix

\section{Boundary expansions}

Along this paper we used shooting methods to get the solutions to the e.o.m.
We present here the (next to) leading order behavior of the fields near the boundary that is necessary
to carry out the numerics.

For large $y\rightarrow\infty$ the fields admit the expansions,
\bea\label{bdryexp}
A(y)&=& 1 + \frac{A_1}{y^{a_1}}\;\tilde A_1(y)= 1 + \frac{A_1}{y^{a_1}}\;
\left(1 + \frac{A_2}{y^{a_2}}\;\tilde A_2(y)\right)\cr
f(y)&=& y^2 + \frac{F_1}{y}\;\tilde F_1(y)= 1 + \frac{F_1}{y}\;
\left(1 + \frac{\alpha{}^2\,J_1{}^2}{2\,\hat m_W{}^2\,F_1\,y}\;\tilde F_2(y)\right)\cr
c(y)&=& 1 + \frac{C_1}{y^{3}}\;\tilde C_1(y)= 1 + \frac{C_1}{y^3}\;
\left(1 + \frac{C_2}{y^{c_2}}\;\tilde C_2(y)\right)\cr
K(y)&=& \frac{K_1}{y^{\kappa_1}}\;\tilde K_1(y)= \frac{K_1}{y^{\kappa_1}}\;
\left(1 -\frac{J_0{}^2}{2\,(1+2\,\kappa_1)\,y^2}\;\tilde K_2(y)\right)\quad;\quad
\kappa_1\equiv\frac{1}{2} + \sqrt{ \frac{1}{4} +\hat m_W{}^2}\cr
J(y)&=& J_0 + \frac{J_1}{y}\;\tilde J_1(y)= J_0 + \frac{J_1}{y}
\left(1 + \frac{J_2}{y^{j_2}}\;\tilde J_2(y)\right)\cr
H(y)&=& 1 + \frac{H_1}{y^3}\;\tilde H_1(y)= 1 + \frac{H_1}{y^3}\;
\left(1 + \frac{H_2}{y^{h_2}}\;\tilde H_2(y)\right)
\eea
where $\;\tilde A_i(0)=1,\; a_i>0$, etc., for $i=1,2,\dots$.
The constants $(F_1, C_1, J_0, J_1, K_1, H_1)$ are free, all the other ones as well the powers (including those make explicit in (\ref{bdryexp})) are determined by the e.o.m.
\footnote{
The boundary conditions at the horizon leave just one free parameter (that we take $J_0$) that determines completely the solution, see Section \ref{numersn}.
}
In the isotropic case, $\;K(y)\equiv K_1(y)=K_2(y)$,  $C(y) =1\; (C_1=0)\;$,
they are given by,
\bea
(a_1 , A_1) &=& \left\{\begin{array}{lcccc}
6<2\,\kappa_1+2&,& -\frac{3}{4}\,\alpha{}^2\,H_1{}^2&;&
2<\hat m_W{}^2<\infty\\
2\,\kappa_1+2<6&,& -\frac{\alpha{}^2\,\kappa_1{}^2\,K_1{}^2}{2\,\hat m_W{}^2
(1+\kappa_1)}&;& 0<\hat m_W{}^2<2\\
6=2\,\kappa_1+2&,& -\alpha{}^2\,\left(\frac{3}{4}\,H_1{}^2 +
\frac{1}{3}\,K_1{}^2\right)&;& \hat m_W{}^2=2
\end{array}\right.\cr
(j_2 , J_2) &=& \left\{\begin{array}{lcccc}
6<2\kappa_1+1&,& \frac{A_1}{7}&;& \frac{15}{4}<\hat m_W{}^2<\infty\\
2\kappa_1+1<6&,& \frac{K_1{}^2\,J_0}{(2\,\kappa_1+1)\,(1+\kappa_1)\,J_1}
&;& 0<\hat m_W{}^2<\frac{15}{4}\\
6=2\,\kappa_1+1&,& \frac{K_1{}^2\,J_0}{21\,J_1}+\frac{A_1}{7}&;& \hat m_W{}^2=\frac{15}{4}\end{array}\right.\cr
(h_2 , H_2) &=& \left\{\begin{array}{lcccc}
3<2\kappa_1 -1&,& -\frac{F_1}{2}&;& 2<\hat m_W{}^2<\infty\\
2\kappa_1 -1<3&,& \frac{K_1{}^2}{(2\,\kappa_1-1)\,(1+\kappa_1)\,H_1}
&;& 0<\hat m_W{}^2<2\\
3=2\,\kappa_1-1&,& \frac{K_1{}^2}{9\,H_1}-\frac{F_1}{2}&;&\hat m_W{}^2=2
\end{array}\right.
\eea
while that in the anisotropic case, $\;K(y)\equiv K_1(y), \;K_2(y)=0$, they result,
\bea
(a_1, A_1) &=&(3, - C_1)\\
(c_2 , C_2) &=& \left\{\begin{array}{lcccc}
3<2\kappa_1 -1&,& -\frac{F_1-C_1}{2}&;& 2<\hat m_W{}^2<\infty\\
2\kappa_1 -1<3&,& -\frac{\alpha{}^2\,(1+\hat m_W{}^{-2}\,\kappa_1{}^2)\,K_1{}^2}{2\,
(2\,\kappa_1-1)\,(1+\kappa_1)\,C_1}&;& 0<\hat m_W{}^2<2\\
3=2\,\kappa_1-1&,& -\frac{F_1-C_1}{2} -\frac{\alpha{}^2\,K_1{}^2}{6\,C_1}&;&
\hat m_W{}^2=2
\end{array}\right.\cr
(j_2, J_2 ) &=&\left(3,-\frac{C_1}{2}\right)\cr
(h_2 , H_2) &=& \left\{\begin{array}{lcccc}
3<2\kappa_1 -1&,& -\frac{F_1}{2}&;& 2<\hat m_W{}^2<\infty\\
2\kappa_1 -1<3&,& \frac{K_1{}^2}{2\,(2\,\kappa_1-1)\,(1+\kappa_1)\,H_1}
&;& 0<\hat m_W{}^2<2\\
3=2\,\kappa_1-1&,& -\frac{F_1}{2}+\frac{K_1{}^2}{18\,H_1}&;&
\hat m_W{}^2=2
\end{array}\right.
\eea

\section{AH conventions}

In this appendix we write the e.o.m. in the conventions of references \cite{Ammon:2009xh}, \cite{HHV}.
Besides being often present in literature, they proved to be convenient in some numerical computations.

The ansatz is written as,
\bea\label{gralansatz2}
g &=& \frac{L^2}{u^2}\;\left(- \tilde f(u)\; s(u)^2\; {dt}^2 + \frac{dx^2}{g(u)^2} + g(u)^2\,{d\tilde y}^2
+ \frac{du^2}{\tilde f(u)}\right)\cr
A &=& dt\;\tilde J(u)\;X_0+ dx\;\tilde K_1(u)\;X_1 + d\tilde y\;\tilde K_2(u)\;X_2\cr
H &=& H_0\;\tilde H(u)\; X_0
\eea
The relation with the conventions used in the bulk of the paper are,
\bea\label{relconHHV}
&&x^0 =L\,t\quad,\quad x^1 =L\,x\quad,\quad x^2 =L\,\tilde y\quad,\quad y= \frac{g(u)}{u}\cr
&&c(y)= \frac{1}{g(u)^2}\quad,\quad  f(y)= \frac{\tilde f(u)}{u^2}\,(g(u)-u\,g'(u))^2 \quad,\quad
A(y)= \frac{s(u)}{g(u)-u\,g'(u)}\cr
&&J(y)=\tilde J(u)\quad,\quad K_1(y)= K_1(u)\quad,\quad K_2(y)=\tilde K_2(u)\quad,\quad
H(y) = \tilde H(u)
\eea
The gravity  e.o.m. result (out the tildes),
\bea\label{graeom2}
f'(u) &=& \frac{3}{u}\,(f(u)-1) + u\,f(u)\,\frac{g'(u)^2}{g(u)^2} + \frac{u}{g(u)^2}\; (1)\cr
\frac{s'(u)}{s(u)} &=& - u\,\frac{g'(u)^2}{g(u)^2} - \frac{u^3}{g(u)^2}\;y'(u)^2\,(2)\cr
2\,\frac{u^2\,g(u)}{f(u)\,s(u)}\;\left( \frac{s(u)\,f(u)}{u^2\,g(u)}\,g'(u)\right)' &=& -\frac{1}{f(u)\,g(u)}\;(3)
\eea
where $(1), (2), (3)$ are the r.h.s.'s of (\ref{graeom1}) written in the variables (\ref{relconHHV}),
while that the matter e.o.m. are,
\bea\label{matteom2}
\frac{1}{g(u)^2}\,\left(f(u)\,s(u)\,g(u)^2\, K_1'(u) \right)'
&=& \left( \frac{s(u)}{g(u)^2} \,K_2(u)^2+\hat m_W{}^2\; \frac{s(u)\,H(u)^2}{u^2}  -
\frac{J(u)^2}{f(u)\, s(u)} \right)\;K_1(u)\cr
\frac{1}{g(u)^2}\,\left(f(u)\,s(u)\,g(u)^2\, K_2'(u) \right)'
&=& \left( s(u)\,g(u)^2 \,K_1(u)^2+\hat m_W{}^2\,\frac{s(u)\,H(u)^2}{u^2}  -
\frac{J(u)^2}{f(u)\, s(u)} \right)K_2(u)\cr
f(u)\,s(u)\,\left( \frac{J'(u)}{s(u)}\right)' &=& \left(g(u)^2\,K_1(u)^2+ \frac{K_2(y)^2}{g(u)^2}\right)\;J(u)\cr
\frac{u^2}{s(u)}\,\left( \frac{f(u)\,s(u)}{u^2}\, H'(u)\right)' &=& \left( g(u)^2\,K_1(u)^2+ \frac{K_2(y)^2}{g(u)^2}
+ \frac{\lambda_0}{\hat m_W{}^2}\,\frac{H(u)^2-1}{u^2}\right)\;H(u)\cr
& &
\eea
Finally, the contributions to the free energy density are,
\bea\label{f2}
\hat f^{(bulk)} &=& \frac{\alpha^2}{2\,J_0{}^3}\,\int^{u_1}_{u_0}\,du\,s(u)\,
\left(\frac{6}{\alpha{}^2\,u^4} -\frac{\lambda_0}{2\,u^4}\,\left(H(u)^2-1\right)^2\right.\cr
&+&\frac{f(u)}{g(u)^2}\,\left(g(u)^4\,K'_1(u)^2 + K'_2(u)^2\right)
+K_1(u)^2\,K_2(u)^2\cr
&-&\left.\frac{J'(u)^2}{s(u)^2}-\frac{J(u)^2}{f(u)\,s(u)^2\,g(u)^2}\,\left(g(u)^4\,K_1(y)^2 + K_2(y)^2\right)
\right)\cr
\hat f^{(GH)} &=& \frac{\alpha^2}{2\,J_0{}^3}\,\frac{2}{\alpha{}^2}\,u\,f(u)^\frac{1}{2}\,
\left(\frac{f(u)^\frac{1}{2}\,s(u)}{u^3}\right)'|_{u_0}\cr
\hat f^{(ct)} &=& \frac{\alpha^2}{2\,J_0{}^3}\,\frac{4}{\alpha^2}\,
\left(\frac{f(u)^\frac{1}{2}\,s(u)}{u^3}\right)|_{u_0}
\eea

\end{document}